\documentclass[useAMS,usenatbib]{mn2e}
\pdfoutput=1
\usepackage{graphicx}
\usepackage{longtable} 
\usepackage{deluxetable}
\usepackage{lscape}

\graphicspath{{plots/}}

\def\apjl{ApJL }
\def\aj{AJ }
\def\apj{ApJ }
\def\pasp{PASP }

\def\apjs{ApJS }

\def\aap{A\&A }
\def\aapr{A\&AR }
\def\aaps{A\&AS }

\def\mnras{MNRAS }
\def\apss{Ap\&SS }
\def\bain{BAIN }

\def\halpha{\mbox{H$\rm \alpha$}}

\def\sloanr{\mbox{r'}}
\def\sloani{\mbox{i'}}
\def\ri{\mbox{(r'-i')}}
\def\rha{\mbox{(r'-$\rm {H\alpha}$)}}
\def\ewha{\mbox{EW$\rm _{H\alpha}$}}
\def\lacc{\mbox{L$\rm _{acc}$}}
\def\lha{\mbox{L$\rm _{H\alpha}$}}
\def\macc{\mbox{$\rm \dot M_{acc}$}}
\def\mass{\mbox{$\rm M_\star$}}
\def\radius{\mbox{$\rm R_\star$}}
\def\lum{\mbox{$\rm L_\star$}}
\def\rin{\mbox{R$\rm _{in}$}}

\def\msol{\mbox{$\rm M_\odot$}}

\def\sqdeg{\mbox{${\rm deg}^2$}}
\def\deg{\hbox{$^\circ$}}

\def\lesssim{\mathrel{\hbox{\rlap{\hbox{\lower4pt\hbox{$\sim$}}}\hbox{$<$}}}}
\def\gtrsim{\mathrel{\hbox{\rlap{\hbox{\lower4pt\hbox{$\sim$}}}\hbox{$>$}}}}

\begin{document}
\title[T Tauri candidates in IC\,1396 using IPHAS]{
T Tauri candidates and accretion rates using IPHAS:
method and application to IC\,1396}
\author[Barentsen et al.]
{Geert Barentsen$^1$\thanks{E-mail: gba@arm.ac.uk}, Jorick S. Vink$^1$, J. E. Drew$^2$, R. Greimel$^3$, N. J. Wright$^4$, \newauthor
J. J. Drake$^4$, E. L. Martin$^5$, L. Valdivielso$^6$, R. L. M. Corradi$^{6,7}$ \\
$^1$Armagh Observatory, College Hill, Armagh BT61 9DG, U.K.\\
$^2$Centre for Astronomy Research, Science and Technology Research Institute, University of Hertfordshire, Hatfield AL10 9AB, U.K.\\
$^3$Institut f\"ur Physik, Karl-Franzens Universit\"at Graz, Universit\"atsplatz 5, 8010 Graz, Austria\\
$^4$Harvard-Smithsonian Center for Astrophysics, MS-67, 60 Garden Street, Cambridge, MA 02138, USA\\
$^5$Centro de Astrobiolog\'ia (CSIC/INTA), 28850 Torrej\'on de Ardoz, Madrid, Spain\\
$^6$Instituto de Astrof\'\i sica de Canarias, E-38200 La Laguna, Tenerife, Spain\\
$^7$Departamento de Astrof\'\i sica, Universidad de La Laguna, E-38206 La Laguna, Tenerife, Spain
}

\date{Received 19 January 2011; Accepted 8 March 2011}

\pagerange{\pageref{firstpage}--\pageref{lastpage}} \pubyear{2011}

\maketitle

\label{firstpage}

\begin{abstract}
	The INT Photometric H-Alpha Survey (IPHAS) is a 1800 deg$^2$ survey of the Northern Galactic Plane, reaching down to r' $\sim$21. 
We demonstrate how the survey can be used to 
(1) reliably select classical T Tauri star candidates and 
(2) constrain the mass accretion rates with an estimated relative uncertainty of 0.6 dex. 
IPHAS is a necessary addition to spectroscopic surveys because it allows large 
and uniform samples of accretion rates to be obtained with a precise handle on the selection effects.

	We apply the method on a region of 7 deg$^2$ towards the H{\sc ii} region IC\,1396 in Cepheus OB2 
and identify 158 pre-main sequence candidates with masses between 0.2 and 2.0 \msol\ and accretion rates between $10^{-9.2}$ and $10^{-7.0}~M_\odot yr^{-1}$.
We find a power-law dependency between the stellar mass and the accretion rates with a slope of $\alpha~=~1.1\pm0.2$,
which is less steep than indicated by previous studies. 
We discuss the influence of method-dependent systematic effects on the determination of this relationship.
 
	The majority of our sample consists of faint, previously unknown, low-mass T Tauri candidates 
(56 per cent between 0.2 and 0.5 $M_\odot$).  
Many candidates are clustered in front of three bright-rimmed molecular clouds,
which are being ionized by the massive star HD\,206267 (O6.5V).
We discover a spatio-temporal gradient of increasing accretion rates,
increasing Spitzer infrared excess, and younger ages away from the ionizing star,
providing a strong indication that the formation of these clusters has been sequentially triggered 
by HD\,206267 during the last $\sim$1~Myr.
\end{abstract}


\section{Introduction}
	In the current picture of star formation, young solar-like stars grow by accreting gas from a circumstellar disc \citep{hartmann_book}.
Whilst this picture is widely supported by observations and theory \citep[e.g.][]{appenzeller1989,hartmann1994},
there remain considerable uncertainties with respect to the time-scales and mechanisms 
responsible for the removal of circumstellar disc material \citep[e.g.][]{mayne2008}, 
with fundamental implications for planet-formation models \citep{hester2004,adams2004,throop2005,hollenbach2005,hillenbrand2005}.
One of the main reasons why progress has been slow is that disc evolution is thought to be
driven by a complex interplay between several physical effects, involving 
magnetospheric accretion, UV-photoevaporation, stellar winds, cosmic ray and X-ray ionization, dust coagulation, 
planet formation, and binary interactions \citep{hillenbrand2008}. 

	Another hindrance to progress is the distance of typical star forming regions.
Most studies to date have concentrated on relatively small numbers of objects (several tens) 
in nearby regions (e.g., Taurus, Ophiuchus and Chamaeleon at $\sim$150~pc),
because this is where high-quality spectra can be obtained.
These regions cover only a limited range of stellar ages, and are not necessarily representative 
for the massive OB associations 
where the majority of solar-type stars are thought to form \citep[e.g. Cygnus OB2,][]{wright2010}.
In fact, there is increasing evidence to suggest that our Solar System formed near a massive star \citep[e.g.][]{hester2005}.
Unfortunately nearly all OB regions are located beyond 500~pc, where high-resolution spectroscopy of large numbers 
of stars with solar-like masses becomes increasingly challenging \citep[see e.g.][]{vink2008}. 

	Whilst large and unbiased statistical samples of young stars across star-forming regions are required 
for a significant breakthrough in our understanding of star formation, such samples have 
only recently become available through infrared photometry \citep[e.g.][]{evans2009} which traces the circumstellar {\it dust}. 
However, large samples are not yet available for optical (emission-line) studies 
which traces material in the {\it gas} phase, which dominates the circumstellar environment.

	Over the past decade, gas accretion rates have been derived from UV and optical continuum emission 
produced by the impact shock \citep[e.g.][]{gullbring1998,herczeg2008},
as well as the shape and intensity of emission lines produced in the accreting gas \citep[e.g.][]{hartigan2003,natta2004,mohanty2005,dahm2008}.
The available measurements do not provide a particularly clear picture: at any given age, stars 
are observed with widely different accretion rates and disc properties 
\citep{calvet2000,muzerolle2003,sicilia2006spitzer,sicilia2010,currie2009}.
For example, whilst roughly 50 per cent of stars appear to end accretion and lose their inner discs by the age of 1~Myr, 
some small fraction of stars appear to continue accreting beyond 10~Myr \citep{haisch2001,armitage2003,mamajek2004}.

	\cite{demarchi2010} recently proposed a strategy to obtain statistical samples of mass accretion rates without the need for spectroscopy.
The authors combined HST broadband V/I photometry with narrow-band \halpha\ imaging 
to select a large number of pre-main sequence candidates in the Large Magellanic Cloud, and to determine 
their mass accretion rates. Although the lack of spectroscopic follow-up might compromise 
the accuracy of their results for individual objects, 
their work produced a large and homogeneous sample which 
allows a sound statistical appraisal to be made.

	In this paper, we introduce a similar approach based on \sloanr/\sloani/\halpha\ photometry obtained from the 
IPHAS survey of the Northern Galactic Plane. In brief, we set-out to determine mass accretion rates 
from the (\sloanr-\sloani/\sloanr-\halpha) colour-colour diagram, 
whilst simultaneously estimating stellar ages and masses from the (\sloanr/\sloanr-\sloani) colour-magnitude diagram. 
Because we consider objects in a star-forming region for which some 
spectroscopic results are already in the literature, we can examine how 
our photometric approach fares in comparison with more conventional 
spectroscopic methods. 

\begin{figure*}
\centering
\includegraphics[width=0.95\textwidth]{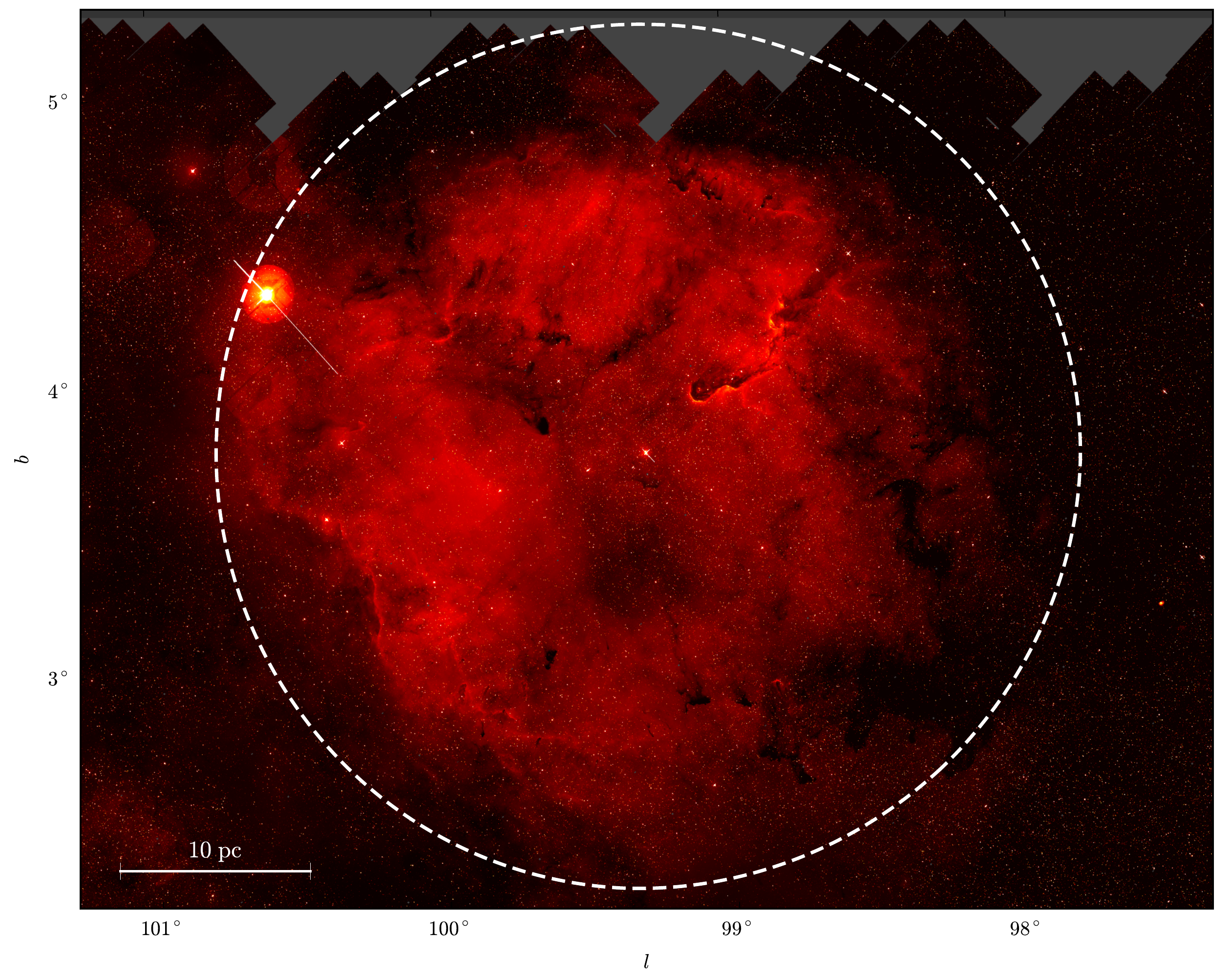}
\caption{\label{fig:mosaic} \halpha-mosaic of IPHAS observations towards IC\,1396 at the edge of Cepheus~OB2. The white dashed line shows the region studied in this paper,
with a radius of 1.5\deg\ centered on the massive star system HD\,206267 (O6.5V), which is the main ionizing source of the region.
Grey areas near the top indicate the upper edge of the IPHAS survey. 
}
\includegraphics[width=0.95\textwidth]{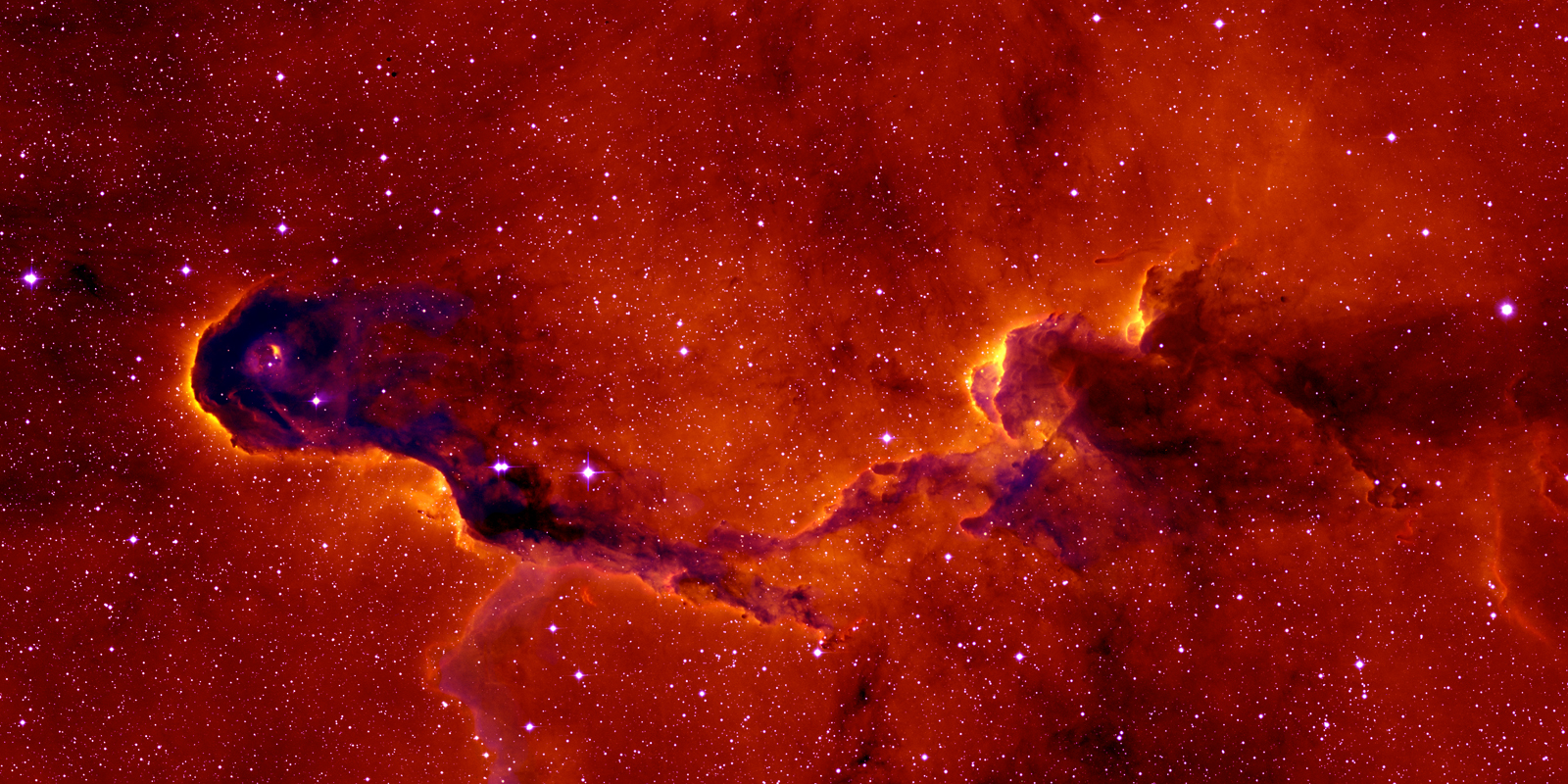}
\caption{\label{fig:nick} Close-up color mosaic of IC1396a ($l=99\deg, b=+4\deg$), commonly known as the Elephant's Trunk Nebula,
composed from the \halpha\ (red channel), \sloanr\ (green) and \sloani\ (blue) bands. 
Red and yellow colors indicate regions with the most intense \halpha\ emission.} 
\end{figure*}

As a testbed, we study the H{\sc ii} region IC\,1396, 
ionized by the O-star HD\,206267.
A comprehensive review of IC\,1396 and its associated open cluster Trumpler~37 is given in \citet{kun2008}.
In brief, the region is located at the edge of the Cepheus OB2 association at a distance of $\sim$900~pc \citep{contreras}.
It has previously been the target of photographic \halpha\ objective prism surveys \citep{kun1986,kun1990,balazs1996}.
However, these surveys were not deep enough (R $\lesssim$ 15-16) to uncover the low-mass T Tauri population (R $>$ 16),
because slit-less spectroscopy becomes increasingly impractical at faint magnitudes in crowded fields.
The IPHAS digital narrow-band imaging is better suited for this purpose.

The central part of IC\,1396 has been the target of an optical survey by \citet{sicilia2004,sicilia2005,sicilia2006,sicilia2006spitzer,sicilia2010},
who uncovered a rich population of T Tauri stars using photometric candidate selection combined with follow-up spectroscopy.
Our work extends this existing survey, as our photometric approach allows us to probe 2 magnitudes deeper and study an area that 
is a factor four times larger (including some of the unexplored outskirts of IC\,1396).

We identify 158 accreting pre-main sequence candidates from a database of $\sim$1 million objects,
with a contamination rate of less than 15 per cent and well-defined detectability limits.
The sample allows us to investigate the dependency of accretion rates on stellar age and mass,
and shows a spatio-temporal gradient which, we will argue, provides evidence for 
sequentially triggered star formation.

The paper is organized as follows:
in \S 2 we introduce the IPHAS survey and explain how T Tauri candidates are selected from the colour-colour diagram.
In \S 3 we show how the ages, masses and accretion rates are determined,
and in \S 4 we explain the reduction of archival infrared photometry.
In \S 5 we present the results 
and in \S 6 we use existing literature to show that the vast majority of our candidates are genuine T Tauri stars.
A careful analysis of the errors is presented in \S 7.
Finally, in \S 8 we discuss the implications of our results and \S 9 contains the conclusions.

\section{Method: Selecting T Tauri candidates using IPHAS colours}
\subsection{IPHAS observations of IC 1396}
\label{sec:observations}
A comprehensive overview of the IPHAS survey is given in~\citet{drew2005,idr} and on the project website\footnote{http://www.iphas.org}. 
Briefly, IPHAS is a 1800 \sqdeg\ photometric survey of the Northern Galactic Plane 
($30^{\rm o} \lesssim \ell \lesssim 220^{\rm o}$, $-5^{\rm o} \lesssim b \lesssim +5^{\rm o}$), 
carried out using the Wide Field Camera (WFC) on the 2.5-meter Isaac Newton Telescope (INT) in La Palma. 
Photometry was obtained using a narrow-band \halpha\ filter and the broadband Sloan \sloanr\ and \sloani\ filters. 
All data are then pipeline processed at the Cambridge Astronomical Survey Unit~\citep[CASU;][]{irwin1985,irwin}.
This routinely includes photometric calibration based on nightly observations 
of standard star fields.  However we have been able to draw upon the 
results of a first-pass global calibration of IPHAS survey data (Miszalski 
et al, in preparation), 
thereby significantly reducing field-to-field magnitude shifts.

The observations of IC\,1396 were obtained during several bright but mostly photometric nights in August 2004. 
Fields that suffered from nights of poor seeing were repeated in subsequent years (2005-2009), resulting in a final average seeing of $0.9 \pm 0.1"$. 
We used the {\sc Montage}\footnote{http://montage.ipac.caltech.edu} toolkit to create an \halpha\ mosaic of the 120 IPHAS fields towards the region, 
shown in Fig.~\ref{fig:mosaic}. A close-up of the Elephant's Trunk Nebula, which is part of the region, is shown in Fig.~\ref{fig:nick}.

The contrast of the mosaic has been stretched to bring out the strong H$\alpha$ background emission in the region.
We draw the reader's attention to the ability of the CASU pipeline to track variations in the background emission on scales of 20-30 arcsec. 
This allows the stellar flux to be accurately separated from the background in the vast majority of cases. 
Any objects which, nevertheless, suffer from incorrect background subtraction due to very rapidly spatially varying regions of high nebulosity
are flagged as part of the pipeline's morphological classification. 
In brief, the pipeline derives a curve-of-growth for each detected object from a series of aperture flux measures with different aperture radii. 
In the case of incorrect background correction, the curve-of-growth will not show the 
asymptotic behaviour expected of stellar sources and so will be classified as non-stellar.
We confirmed the effectiveness of this feature using manual inspection of pipeline photometry 
in multiple nebulous regions.

\subsection{Quality criteria}
\label{sec:quality}
The IPHAS database contains more than 1 million photometric detections in our region of study 
(indicated by the dashed circle in Fig.~\ref{fig:mosaic}).
We narrowed down the catalogue to 404\,975 sources using the following strict quality requirements:
\begin{enumerate}
 \item the magnitude must be in the range $13\,<\,r'\,<\,20$;
 \item the star must be morphologically classified as strictly stellar in all bands (``class = -1'');
 \item the difference in coordinates between the matched detections in the \halpha, \sloanr\ and \sloani\ images must not be bigger than $0.1"$.
\end{enumerate}
Criterion (i) avoids faint and saturated sources. Although the IPHAS photometry goes down to a $3\sigma$-depth of $r'=21\pm1$, 
we choose to limit our study to $r' < 20.0$ to keep the photometric uncertainties small and to avoid spurious detections.

Criterion (ii) avoids saturated and extended objects or problems resulting from flux-contamination by nearby neighbours. 
The criterion also deals with the issue of nebulous \halpha\ background emission. 
As noted in the previous section, background-contaminated photometry will be flagged as non-stellar and removed this way.
We note that this has the effect of removing a small number of interesting -- but unreliable -- objects.

Criterion (iii) avoids spurious mismatches in crowded fields. 
The CASU pipeline allows matches up to 2.5 arcseconds apart between different filters. 
This criterion needs to be tightened considerably to avoid mismatches near the faint limit of the survey.

\subsection{Quantifying the \halpha\ excess from IPHAS colours}
\label{sec:colours}

\subsubsection{Colour simulations}
As described by~\citet{drew2005}, the \rha\ colour of a point 
source is an indicator of the strength of the \halpha\ line relative to 
the red continuum.
This may flag the object as an emission line star or, for most stars, 
provide a measure of intrinsic colour.
The \ri\ colour reflects a combination of interstellar reddening 
and intrinsic colour.

Stars with \halpha\ in emission show higher \rha\ values than other stars of a similar spectral type.
However, there is no straightforward analytical relationship between the \rha\ colour and the strength of the \halpha\ line,
in part because the \halpha\ band falls inside the \sloanr\ band.
A given amount of \halpha\ excess emission will affect both magnitudes in a way that depends on the underlying SED and reddening of the star.

For the purpose of estimating the \halpha\ excess from IPHAS data,
we created a set of simulated colour tracks for stars with increasing levels of \halpha\ emission equivalent width (EW) and different underlying SEDs.
A similar grid was presented by \citet[][table 4]{drew2005} 
but did not extend down in effective temperature to stars of K and M type (common among T Tauri stars).

To simulate the colours we adopted the same procedure as described in detail by Drew et al.
In brief, colours are generated from the library of stellar SEDs due to \citet{pickles}, using a standard extinction law $\rm R_{V} = 3.1$ in the form given by \citet{howarth1983},
and taking into account the relevant filter transmission profiles and the CCD response curve available from the WFC website\footnote{http://www.ing.iac.es/Astronomy/instruments/wfc/}.
We adopted as our reference Vega spectrum that due to \citet{bohlin2007}.
The resulting grid is tabulated in Appendix~\ref{app:tracks} and plotted in Fig.~\ref{fig:tracks}.

\begin{figure}
\includegraphics[width=\linewidth]{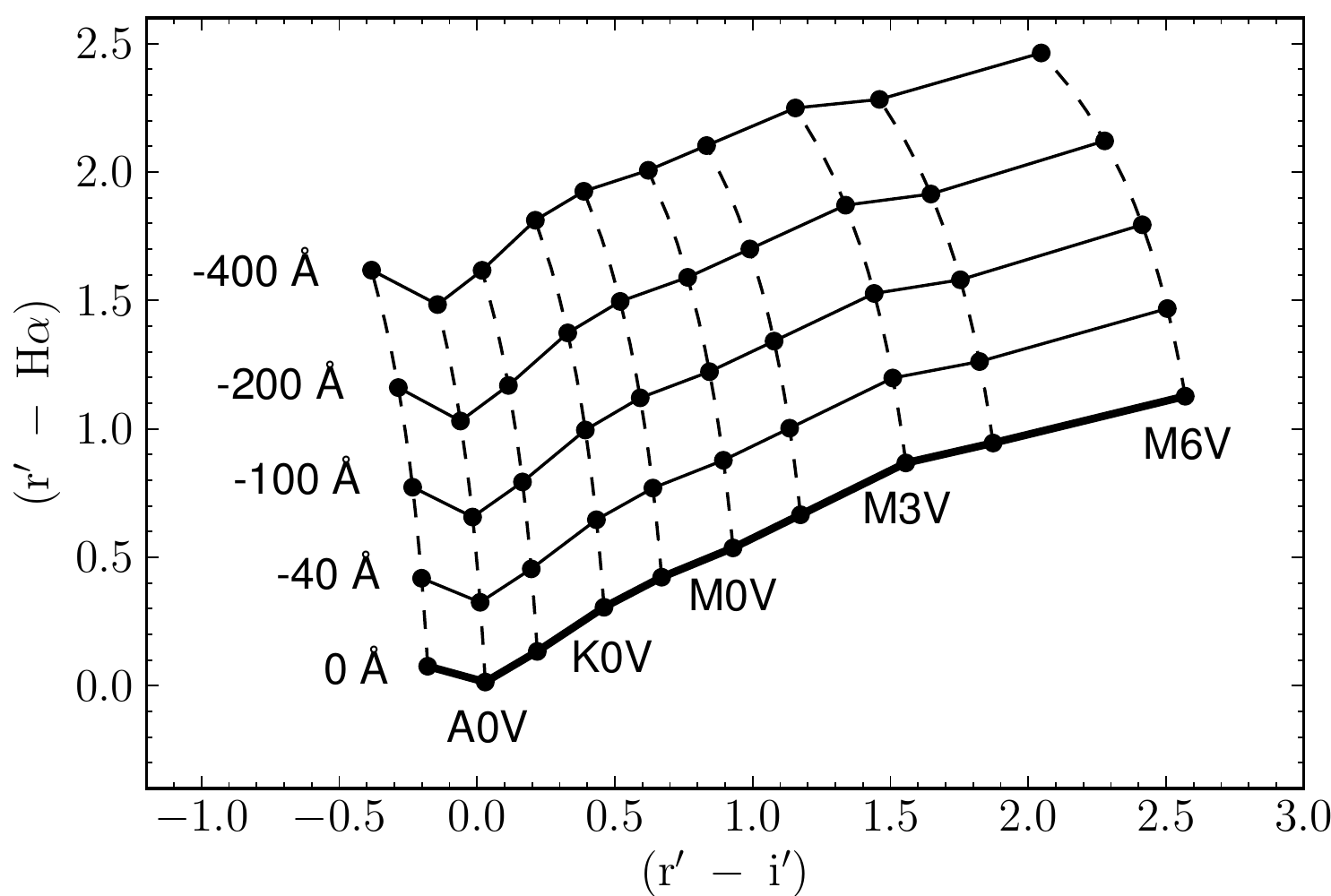}
\caption{Simulated colours for E(B-V)=0 (unreddened stars) with increasing levels of \halpha\ emission. 
The tracks are tabulated for different amounts of reddening in Appendix~\ref{app:tracks}.}
\label{fig:tracks} 
\end{figure}

\subsubsection{Validation of the simulated colours against spectroscopic observations of T Tauri stars}

\begin{figure}
\includegraphics[width=\linewidth]{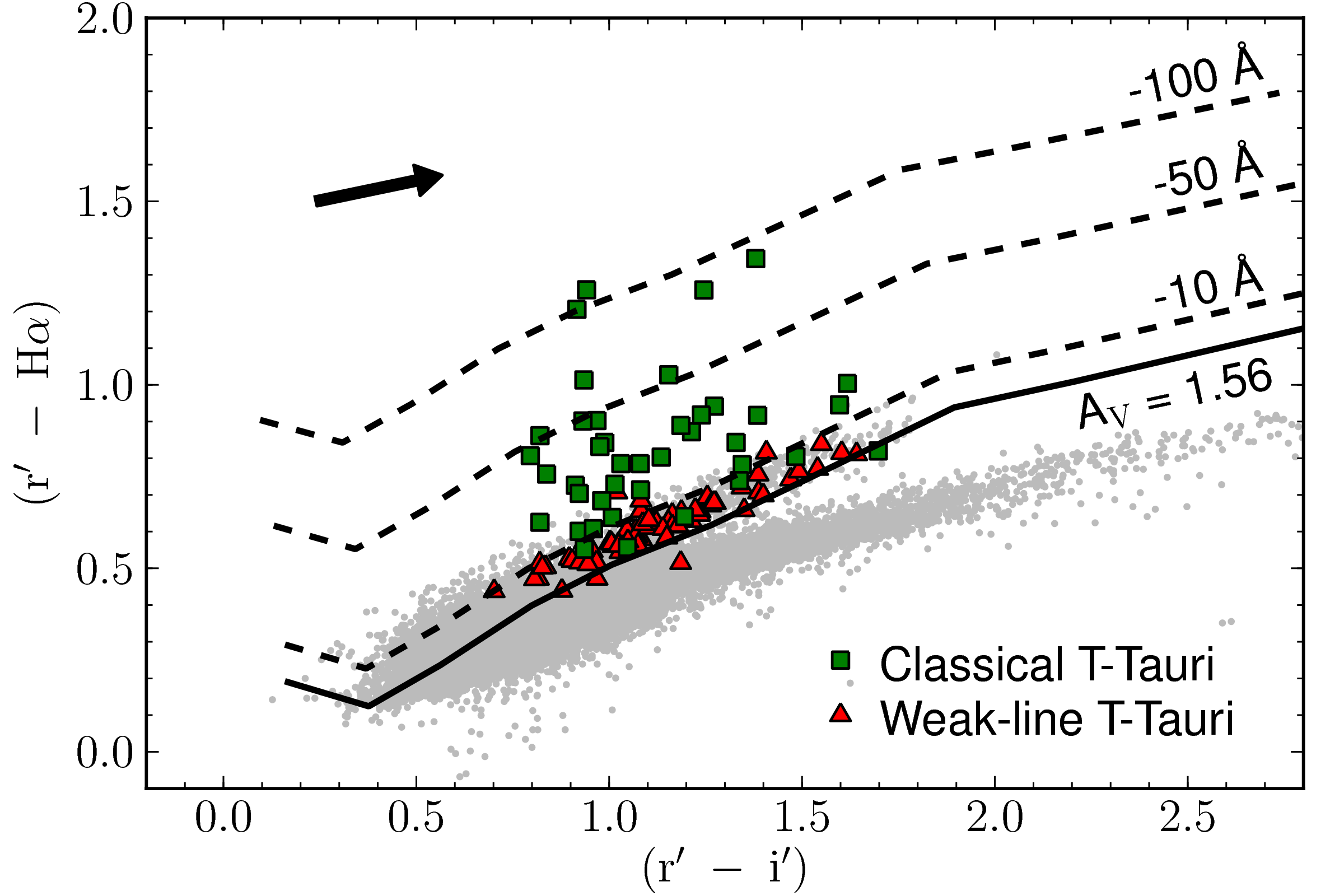}
\caption{IPHAS colours of known T Tauri stars in IC\,1396 from~\protect\citet{sicilia2005}.
Green squares are classical T Tauri stars (CTTS) with spectroscopic \ewha\ stronger than -10\,\AA. 
Red triangles are weak-line T Tauri stars (WTTS) with \ewha\ weaker than -10\,\AA. 
The solid line shows the simulated main sequence curve at the mean reddening of the cluster ($\bar{A}_V=1.56$). 
Dashed lines shows the position of stars at increasing levels of \halpha\ emission as predicted by our simulated tracks.
Grey dots show field stars in the region.
The arrow shows the reddening shift for an M0V-type object being reddened from $\rm A_V=0$ to $\rm A_V=1.56$ 
\citep[note that the true reddening tracks are curved in a way that depends on the SED and the amount of reddening, see][]{drew2005}
}
\label{fig:ccd_aguilar} 
\includegraphics[width=\linewidth]{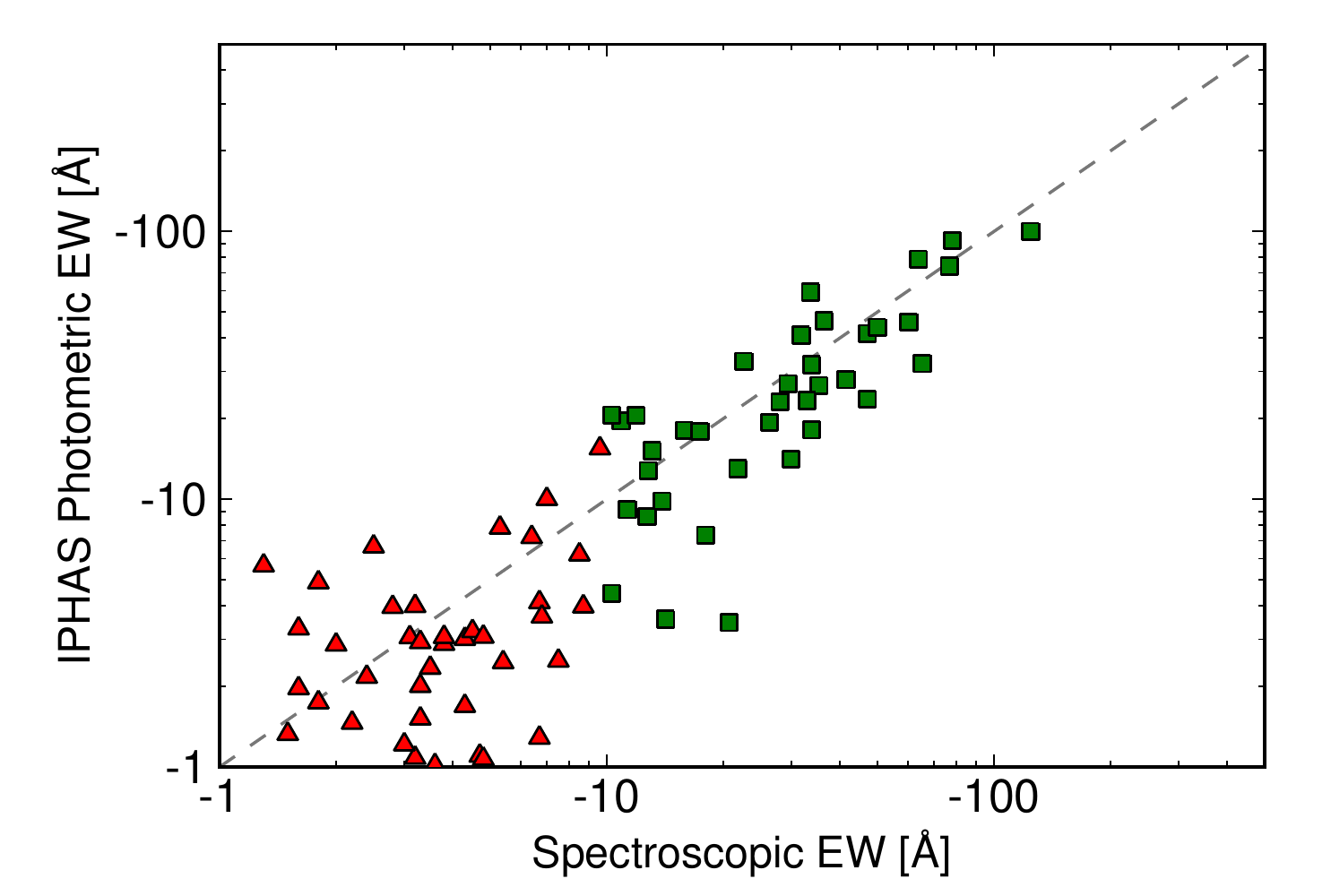}
\caption{Comparison of IPHAS photometric \ewha\ with spectroscopic values from \citet{sicilia2005}.
The grey dashed line shows the unity relation. 
The scatter is thought to be dominated by natural \halpha\ emission variability.}
\label{fig:ew_valid} 
\end{figure}

To test our grid of simulated colours,
we took the sample of known T Tauri stars in IC\,1396 from~\citet[][hereafter referred to as SA05]{sicilia2005}.
In their work, the authors used MMT/Hectospec multifiber spectroscopy 
to follow-up candidates selected using multi-epoch broadband photometry.
The sample includes both weak-line T Tauri stars (WTTS) with \halpha\ emission weaker than -10~\AA\ EW, 
as well as classical T Tauri stars (CTTS) with emission stronger than -10~\AA~EW.

Out of the 118 probable T~Tauri members confirmed by Hectospec observations,
109 have a reliable counterpart in the IPHAS database.
(The other 9 objects where classified as extended objects by the pipeline;
6 objects because they appear to suffer from flux contamination due to a very close neighbour, 
while 3 objects appear to suffer from a rapidly spatially varying background.)

In Fig.~\ref{fig:ccd_aguilar} we show the IPHAS colours of these 109 objects superimposed on the 
grid of simulated lines of constant H$\alpha$ EW 
referred to an appropriately reddened main sequence (zero line emission at 
$A_V = 1.56 \pm 0.55$, which is the typical reddening found by SA05). We notice 
that the classical T~Tauri stars are well separated from the field stars 
(shown as grey points): most are above the $-10$~\AA\ EW 
boundary as predicted by the drawn grid lines.  In contrast the 
weak-lined stars fall within the main stellar locus, blending in with 
normal less-reddened stars. The fact that reddening raises the EW threshold for 
the clean detection of emission line stars is a recognised property of the 
IPHAS colour-colour plane \citep[see][]{drew2005}.

One weak-lined object, named 73-537 in SA05, can be seen to fall somewhat below the simulated main sequence (solid line in Fig.~\ref{fig:ccd_aguilar}).
The aberrant position is explained by the high reddening of the object, $A_V$ = 3.3, 
which is an outlier in terms of reddening compared to the rest of the sample.

To validate the grid in more detail, 
we interpolated the tracks to derive \halpha\ EWs for the known T Tauri objects.
These values are then plotted against the spectroscopic values from SA05.
The comparison is shown in Fig.~\ref{fig:ew_valid}.
We find a strong correlation between the photometric and spectroscopic estimates, albeit with a large scatter on the order of 5 to 10\,\AA.

The scatter is explained by two reasons. 
First, photometric errors and reddening deviations will introduce large relative errors for weak-lined objects which are positioned close to the reddened main sequence line. 
This dominates the scatter observed for objects with lines weaker than -10~\AA. 
Second, net H$\alpha$ emission in T~Tauri stars is known to 
show large natural variations on timescales as short as days \citep[e.g.][Costigan et al, in preparation]{fernandez1995,alencar2001}.
For a subset of the objects considered here, 
\citet[][hereafter referred to as SA06]{sicilia2006} obtained high-resolution MMT/Hectochelle spectroscopy at a second epoch.
Compared to the Hectospec data, the EWs show a scatter of 
33 per cent for emission EWs exceeding -20~\AA\ (fig.~10 in SA06). 
In our sample we observe a similar scatter of 37 per cent for lines stronger than -20~\AA\, 
which suggests that the spread is dominated by intrinsic variations. 

We confirmed this by recomputing photometric equivalent widths, 
this time taking into account the individual reddening values $A_V$ for each object from SA05, 
and find that the scatter decreased only marginally. 
Knowledge of the individual reddening values does not significantly improve the 
estimated line strengths here,
because the spread in reddening is reasonably small for the T~Tauri objects in IC\,1396 ($\sigma_{A_V}=0.55$, SA05).

We conclude that the simulated grid and IPHAS photometry can be used to identify emission-line objects and constrain their 
\halpha\ excess with a sufficient degree of confidence for the purpose of our work, 
which is to constrain the \halpha\ excess for a very large sample of stars - over a million objects in the field towards IC\,1396 
- in a homogeneous way which is not biased by a pre-selected list of observing targets.
To study a similar amount of objects using traditional spectroscopy would require so much observing time that it is effectively impossible.

\subsection{Selection strategy}
\label{sec:selection}
\subsubsection{T Tauri selection threshold}
\begin{figure*}
\centering
\includegraphics[angle=0,width=\linewidth]{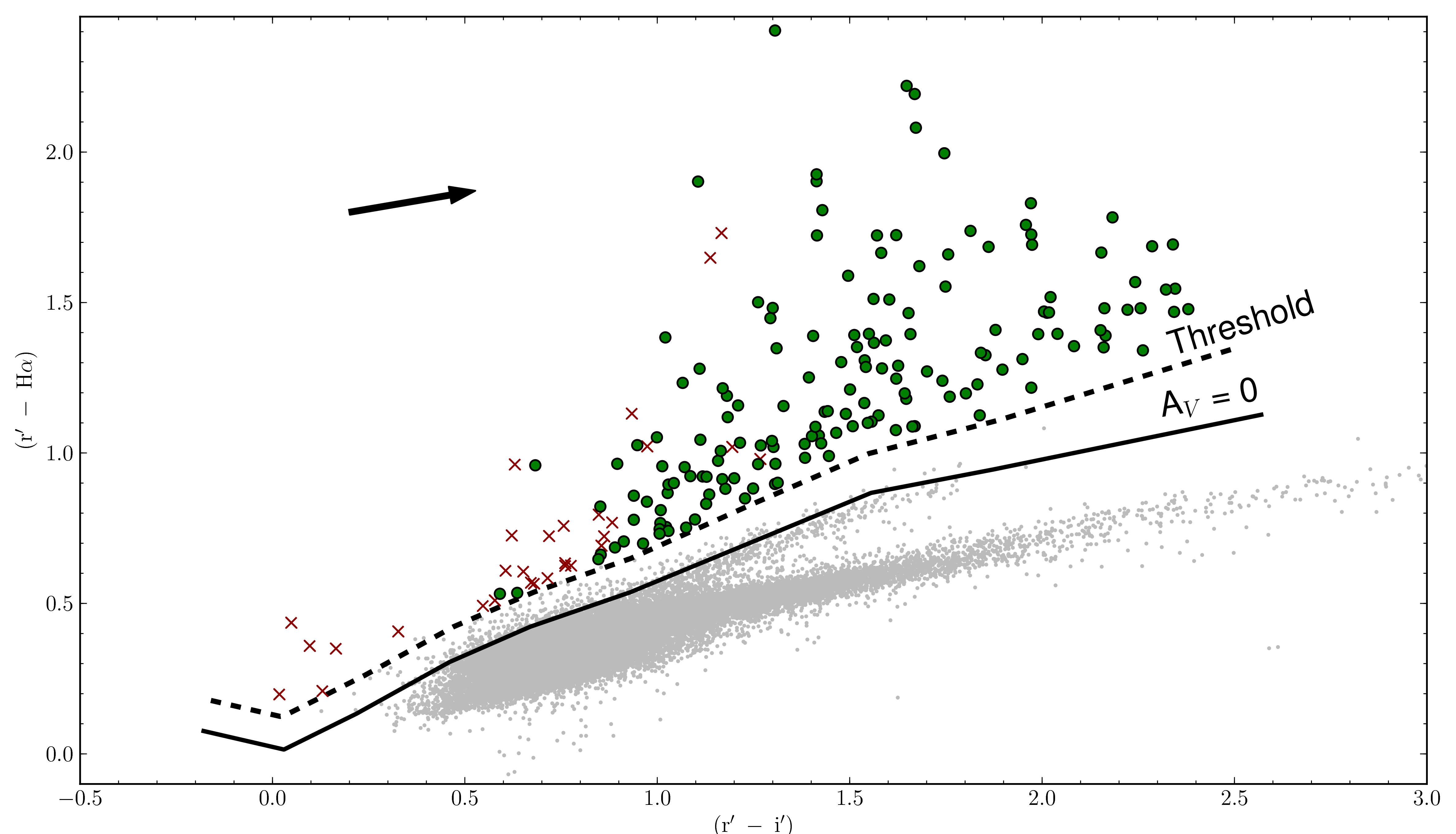}
\caption{\label{fig:ccd_selection} Results of the T Tauri candidate selection towards IC\,1396. 
Green circles show accepted candidates.
Red crosses show stars which were rejected on the basis of their position in the (\sloanr / \sloanr-\sloani) colour-magnitude diagram (Fig.~\ref{fig:cmd}). 
The solid line shows the unreddened main sequence curve from our grid of simulated colours, 
while the dashed line shows the selection threshold which serves to remove chromospherically active foreground dwarfs (see text). 
Grey dots show 50\,000 random fields stars having accurate photometry ($\sigma_{r'}$ $<$ 0.01). 
The arrow shows the reddening shift for an M0V-type object being reddened from $\rm A_V=0$ to $\rm A_V=1.56$.} 
\includegraphics[width=\linewidth]{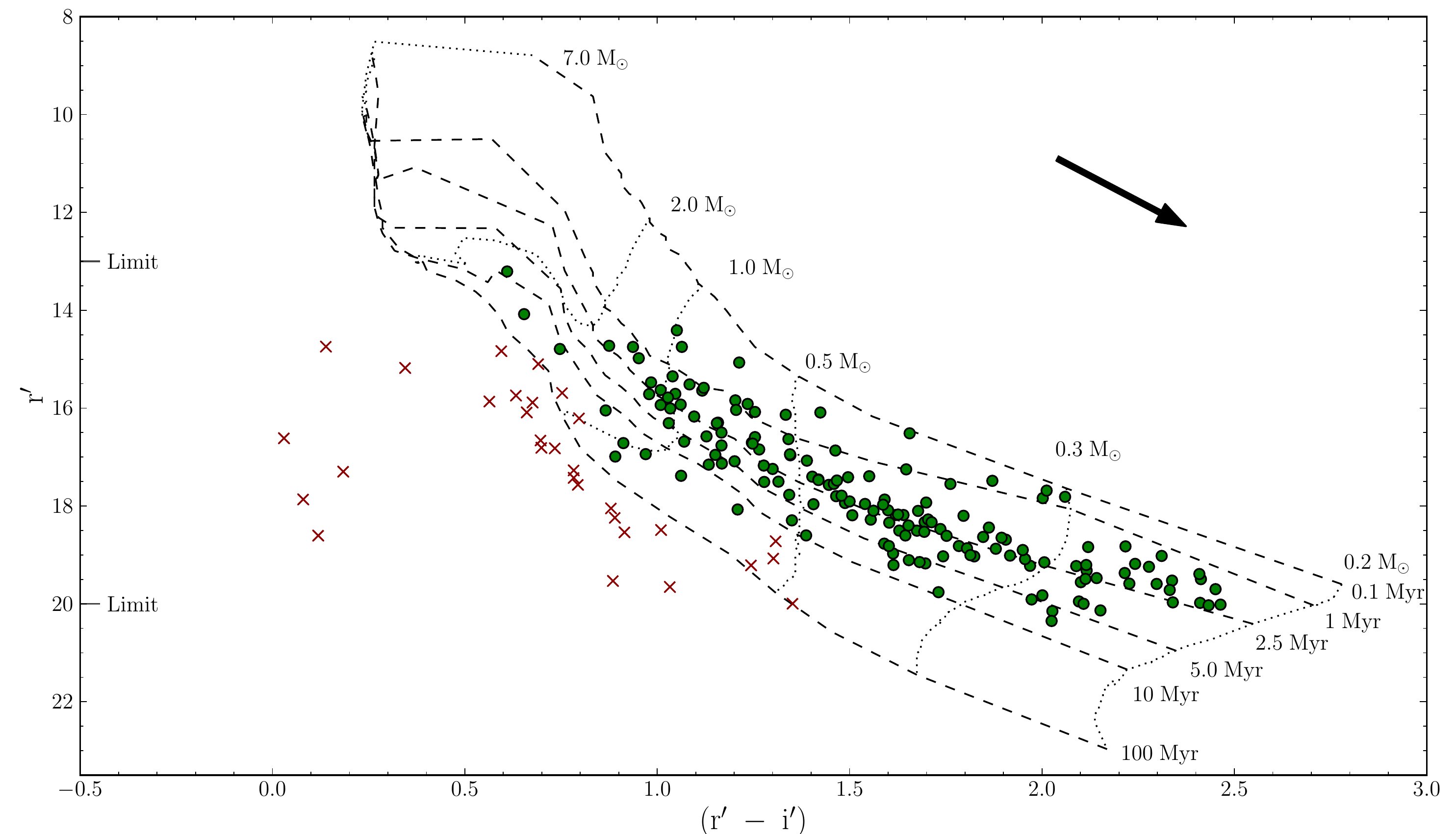}
\caption{\label{fig:cmd} Position of the accepted (green circles) and discarded (red crosses) candidates in the colour-magnitude diagram.
Dashed and dotted lines show isochrones and mass tracks from the \protect\citet{siess} model.
The tracks have been placed at the mean reddening and distance modulus of the cluster,
and the stars have been corrected for the \sloanr-band excess due to the \halpha\ line excess.
Objects are discarded if they are not located above the 100~Myr isochrone with $3\sigma$ confidence (see text).
The arrow shows the reddening vector for $\rm A_V=1.56$ due to \citet{schlegel}. 
}
\end{figure*}
We now make a homogeneous selection of T~Tauri candidates 
based on their position above the main locus of field stars in the (\sloanr-\sloani/\sloanr-\halpha) diagram.
This requires a selection threshold to be established relative to the upper edge of the main locus 
(cf. grey dots in Fig.~\ref{fig:ccd_aguilar}), which is where unreddened foreground main-sequence stars are located.

\defcitealias{barrado}{Barrado y Navascues \& Martin (2003)} 

The threshold is chosen such that the foreground stars are excluded, even when they are chromospherically active.
We adopt the traditional CTTS threshold which requires EWs stronger than -10\,\AA\ for early- and intermediate-type stars.
For late-type stars, which are known to show intense chromospheric activity long after their formation,
we adopt the empirical thresholds found by~\citetalias{barrado} 
ranging from -11\,\AA\ for spectral type M2V (\sloanr-\sloani\ $\,\cong$ 1.2) up to -24\,\AA\ for type M6V (\sloanr-\sloani $\,\cong$\, 2.6).
The resulting threshold is shown as a dashed line in Fig.~\ref{fig:ccd_selection}. 

We stress that the threshold is drawn relative to the \emph{unreddened} main sequence track.
Stars at the typical reddening of IC\,1396 will naturally be located slightly lower in the diagram, 
and will need to show an excess stronger than roughly -25\,\AA\ at all spectral types to fall above the threshold. 
This level of excess is the completeness limit of our work, which is to be discussed in \S \ref{sec:validation}.

We then select all sources which are located above the threshold at the $3\sigma$ confidence level:
\begin{equation}
(r' - H\alpha)_{\rm star} - (r' - H\alpha)_{\rm threshold} > 3\sigma,
\end{equation}
where $\sigma$ is determined from the photometric uncertainty:
\begin{equation}
\sigma = \sqrt{ \sigma^2_{(r' - H\alpha)} + m^2\cdot\sigma^2_{(r' - i')} },
\end{equation}
with $m$ the local slope of the threshold ($\bar{m} = 0.5$).

Fig.~\ref{fig:ccd_selection} shows the result of the selection procedure. 
We find a total of 188 unique objects located confidently above the selection threshold.
In the case of multiple photometric detections due to overlapping or repeated CCD pointings, we averaged the photometry and show the object only once.
We note that no extremely low-mass candidates are found beyond (\sloanr-\sloani) $>$ 2.5,
which is in agreement with the low density of such objects in IPHAS as previously found by \citet{valdivielso2009}.

\subsubsection{Rejecting candidates on the basis of colour-magnitude}
T Tauri stars are not the only \halpha-emitting objects in the Galactic Plane.
Other stellar objects that may show emission include evolved massive stars (e.g., Be stars, Wolf-Rayet stars, luminous blue variables), 
evolved intermediate-mass stars (e.g., Mira variables, unresolved planetary nebulae),
and interacting binaries (e.g., cataclysmic variables, symbiotic stars).
We refer to \cite{corradi2008} for a discussion on the position of these objects in the IPHAS diagram.

Although it is not possible to confirm T Tauri stars solely on the basis of IPHAS photometry, 
these other classes of objects show significantly lower surface densities towards known star-forming regions, 
where faint red young stars have been shown to dominate the local IPHAS diagrams \citep[][figs. 2-3]{witham2008}.
Moreover, the scarce but often luminous nature of these objects implies that they will most likely appear as distant background objects,
at large distance moduli that may push them below the sequence of foreground T Tauri stars in a colour-magnitude diagram. 

This effect is demonstrated in Fig.~\ref{fig:cmd},
where we placed the selected candidates in the (\sloanr/\sloanr-\sloani) diagram.
The objects are superimposed on the model evolutionary tracks and isochrones due to \citet{siess},
which are placed at the mean reddening and distance modulus of the region (discussed in the next section).
A number of blue objects are seen to fall near or below the 100~Myr isochrone (which is a good approximation of the ZAMS for solar mass stars).
Their position cannot be explained by extinction,
because the reddening vector runs nearly parallel to the isochrones.
Instead, these objects are likely to be background objects not related to IC\,1396.

We remove these suspicious objects from our sample by requiring
that all T Tauri candidates are located confidently ($3\sigma$-level) above the 100~Myr isochrone for the region.
From our 188 initial candidates, 30 are discarded on this basis (shown as red crosses).
We note that this criterion, combined with the faint magnitude limit ($r' < 20$),
will not allow us to detect the older population for objects below $\sim$0.4\,\msol, 
and will somewhat bias the survey towards the nearest part of IC\,1396 (where the distance modulus is smaller)

This leaves us with a final sample of 158 T Tauri candidates (green circles),
which are listed in Appendix \ref{tab:candidates} together with their photometry.
In \S \ref{sec:validation} we will show that the majority of these candidates are 
likely to be genuine members because they show spatial clustering and infrared colours consistent with T Tauri objects.

In comparison with the list of confirmed T Tauri members from SA06,
we have 34 objects in common (mostly at bright magnitudes),
while 61 objects were not previously discovered (mostly at faint magnitudes).
The remaining 63 objects fall entirely outside the area that was studied by SA06.
Only 3 objects showing emission stronger than -30\,\AA\ EW in SA06 were not recovered by our selection
(to be discussed in \S \ref{sec:validation}).

In contrast, none of the discarded objects are clustered or appear in the work by SA06.
These objects are listed separately in Appendix \ref{tab:rejected} because they are likely to be interesting emission-line objects, 
but they are not considered in what follows.

\section{Analysis: stellar and accretion properties}
\label{sec:analysis}
Having identified a population of T Tauri candidates, 
we now estimate the ages, masses and accretion rates.
The principal assumption made in this step is that all objects are placed at the mean distance $\bar{d} = 870~\pm~80$~pc~\citep{contreras} 
and mean reddening $\bar{A}_V=1.56~\pm~0.55$~(SA05).

The assumption of a fixed reddening is reasonable because SA05
found that the statistical distribution of $A_V$ for T~Tauri stars in the region is nearly Gaussian,
with $\sim$76 per cent of the objects in the $\pm 1 \sigma$ region, and only three objects -- related to the globules -- showing extinctions beyond $+3\sigma$.
Moreover, SA05 did not find an evident spatial variation of the extinction.
This can be considered consistent with the earlier finding by \citet{morbidelli1997} that most of the reddening towards optical stars in IC\,1396
is fairly uniform and of foreground origin.

Because the distribution of reddening values towards IC\,1396 is constrained,
the uncertainties that follow from the assumption of a fixed reddening are well-understood and will be discussed in \S \ref{sec:errors}.

\subsection{Age, mass and radius}
We estimate ages, masses and radii from the (\sloanr/\sloanr-\sloani) diagram in Fig.~\ref{fig:cmd} using the evolutionary tracks by~\cite{siess}.
We adopt the Siess tracks because they were found to be most consistent in a comparative study between different models by \citet{lawson2001}.
Moreover, these tracks were also adopted in SA05, so by adopting these tracks we are able to compare our results more directly with the existing study.

The Siess model tracks are available for the Cousins $\rm (R_c/R_c-I_c)$ diagram based on temperature-colour conversions due to~\citet{kenyon}.
In the absence of temperature-colour conversions for the Vega-based photometric system used by IPHAS,
we adopt the tracks for Cousins and convert them to (\sloanr/\sloanr-\sloani) using the transformations 
as determined for the WFC by the Cambridge Astronomical Survey Unit\footnote{http://www.ast.cam.ac.uk/$\sim$wfcsur/technical/photom/colours/}:
\begin{equation}
\label{eq:cous1}
\rm r' = R_c + 0.275 \cdot (R_c-I_c) + 0.008 \\
\end{equation}
\begin{equation}
\label{eq:cous2}
\rm (r' - i') = 1.052 \cdot (R_c-I_c) + 0.004
\end{equation}
These transformations are derived from multi-epoch observations of standards due to \citet{landolt1992}.
The measured uncertainty of these transformations ($\sim0.1^m$) 
is several times better than the errors estimated for the position of evolutionary tracks \citep[$0.5^m$ to $1.0^m$,][]{lawson2001}.

Having obtained model tracks in the appropriate system,
we shift the tracks according to the assumed distance and reddening for the region.
Previous observations have shown optical stars in the region to follow a standard extinction law $\rm R_{V} = 3.1$ \citep{morbidelli1997},
allowing us to adopt the reddening relations due to \citet{schlegel}:
\begin{equation}
\rm A_{r'} = 0.843 \times A_V 
\end{equation}
\begin{equation}
\rm A_{i'} = 0.639 \times A_V
\end{equation}
We note that the reddening vector runs nearly parallel to the low-mass isochrones. 
Uncertainties in the extinction will therefore only have a limited effect on the age determination.

As a final step,
an automated routine determines the two closest isochrones and mass tracks for each object and interpolates the model properties using an inverse square distance law. 
The routine also corrects for the \sloanr-band excess due to the \halpha\ line, which falls inside the \sloanr\ band. 
The corrections are based on the table of simulated colours, and range between \sloanr +0.02 (\ewha $\cong$ -20\,\AA) and \sloanr +0.5 (\ewha $\cong$ -500\,\AA).

The resulting ages, masses and radii for all candidates are listed in the Appendix~\ref{tab:properties} and histograms are plotted in Fig.~\ref{fig:hist}.
We note that the colour-magnitude diagram (Fig.~\ref{fig:cmd}) appears to show a large scatter in the ages. 
Whether or not this indicates a true age spread will be discussed in \S \ref{sec:errors}.

\subsection{Mass accretion rate}
\label{sec:macc}
Having determined ages and masses, we now constrain the accretion properties of our candidates.
We will adopt the method introduced by~\citet{demarchi2010}. 

In brief, mass accretion is thought to take place along magnetic field lines 
which act as channels connecting the disc to the star. 
The infalling gas is essentially on a ballistic trajectory, 
falling on to the star at close to free-fall velocities, producing a hot impact shock \citep{calvet1998,gullbring2000}.
The energy released in these shocks heats the infalling circumstellar gas, 
resulting in the broad \halpha\ emission lines seen in classical T Tauri stars.
This implies that the measured line luminosity \lha\ may be used to estimated the accretion luminosity \lacc\ 
and subsequently the mass accretion rate \macc\ \citep[e.g. see][]{muzerolle1998models,kurosawa2006,herczeg2008}.

\subsubsection{Line luminosity (\lha)}
To obtain the line luminosity, \lha, we can safely assume that the \halpha\ line falls entirely inside the breadth of the \halpha\ filter (FWHM = 95\,\AA).

To convert the \halpha\ magnitude and equivalent width into luminosity, 
we recall that IPHAS magnitudes are calibrated in the Vega system.
A specially tailored Kurucz 9400\,K model calibration spectrum of Vega~\citep{bohlin2007}
was used to estimate the average flux of Vega weighted by the \halpha\ throughput curve:
\begin{eqnarray}
\bar{F}_{V}(H\alpha) &=& \frac { \Sigma T(\lambda) F_{V}(\lambda) \Delta\lambda } { \Sigma T(\lambda) \Delta\lambda } \nonumber \\
&=& 1.813 \times 10^{-9}~\mathrm{erg~s^{-1} cm^{-2} \AA^{-1}}
\end{eqnarray}
where $F_{V}$ is the Vega model spectrum
and $T$ is the throughput curve, obtained by multiplying the transmission profile of the \halpha\ filter with the CCD efficiency curve.

This number is then multiplied by the width of the filter
to obtain the appropriate band-integrated reference flux, i.e. $F_0 = 95 \cdot \bar{F}_{V}(H\alpha)$.
This is the flux that corresponds to the magnitude of Vega for which we adopt $m(H\alpha) = 0.03$ 
\citep[for historical and definitional reasons, the magnitude of Vega in the optical is not exactly zero, see][]{bessell1998}.

The flux for any star with a given dereddened magnitude $m(H\alpha)$ is then obtained from:
\begin{equation}
F_{\mathrm band} =  F_0 \cdot 10^{-0.4 \cdot [ m(H\alpha) + 0.03 ]},
\end{equation}
where $m(H\alpha)$ is dereddened using the optical-IR extinction law due to \citet{howarth1983}:
\begin{equation}
A_{H\alpha} = 0.796 \cdot A_V.
\end{equation}

$F_{\mathrm band}$ will contain both the continuum and the line flux,
but we are only interested in the latter.
We may separate the line flux using the equivalent width obtained earlier from the table of simulated colours:
\begin{eqnarray}
F_{\rm line} &=&  F_{\rm band} - F_{\rm continuum} \nonumber \\
&=& F_{\rm band} \cdot \frac{-EW/95}{1 - EW/95},
\end{eqnarray}
which follows from the definition of EW.

Finally, the line luminosity \lha\ is obtained by multiplying the continuum subtracted flux by $4\pi d^2$,
where $d$ is the adopted distance to the region.
We note that \lha\ obtained in this way may include a small contribution from the [N{\sc ii}] forbidden emission lines at 6\,548\,\AA\ and 6\,584\,\AA,
which fall inside the IPHAS \halpha\ filter. 
\cite{demarchi2010} corrected for this effect by assuming that 2.4 per cent of the \halpha\ intensity is due to the [N{\sc ii}] lines.
This correction is based on existing T Tauri line measurements from literature, which show the contribution to be in the range of 0 to $\sim$5 per cent.
Although the effect is very small compared to other sources of error, it appears to be systematic, so we adopt it.

\subsubsection{Accretion luminosity (\lacc)}
\label{sec:lacc}
\begin{figure}
\includegraphics[width=\linewidth]{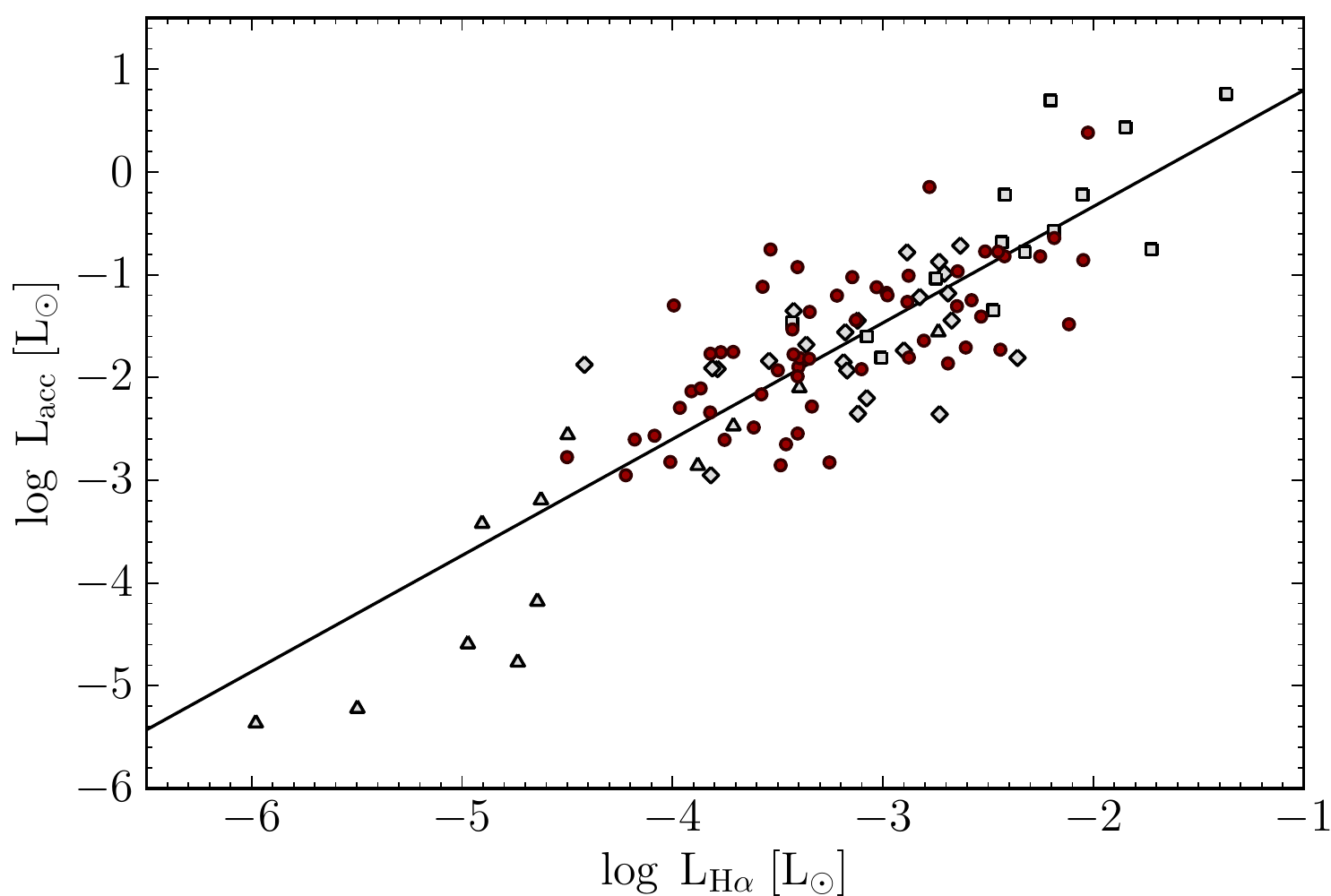}
\caption{\label{fig:lha} Relationship between the \halpha-line luminosity and the accretion continuum luminosity.
Filled circles represent known members of IC\,1396 from \protect\citet{sicilia2010} 
where \lha\ is derived from spectroscopy and \lacc\ is based on U-band photometry. 
We also include literature values from \protect\citep[open diamonds]{hartigan2003}, 
\protect\cite[open triangles]{herczeg2008} 
and \protect\citep[open squares]{dahm2008}. 
The solid line shows the linear fit to all values.}
\end{figure}
As discussed previously, it is generally accepted that the energy \lacc, which is released in the accretion shock, 
goes in part towards heating and ionizing the circumstellar gas in the magnetosphere.
This will cause the gas to reradiate measurable emission lines, 
i.e. we expect a correlation between \lacc\ and \lha.

To study the relationship, 
we compiled a set of T Tauri objects for which measurements of both \lacc\ and \lha\ 
are available in the literature \citep{hartigan2003,herczeg2008,dahm2008}.
In these works, high-resolution spectroscopy was used to measure \lacc\ from the blue continuum excess, 
while \lha\ is derived directly from the \halpha\ line.
After removing lower and upper limit points, we plot the resulting set of 49 objects in Fig.~\ref{fig:lha} as open symbols, showing a clear trend.

We then extend the sample with known members of IC\,1396 
for which \lacc\ was recently determined by \citet{sicilia2010} using deep U-band photometry
and \lha\ being available from their earlier spectroscopic work.
These additional 59 objects are shown as filled red circles.

The data suggests a relationship $(L_{acc}) \propto (L_{H\alpha})^\alpha$ with $\alpha$ between 0.5 and 2.0.
This is consistent with theoretical accretion models due to \cite{muzerolle1998models}.
In their work, the authors find a range of slopes, ranging from $\alpha = 0.7$ (small magnetosphere) to $\alpha = 1.0$ (larger magnetospheres) 
at moderate accretion rates ($\dot M < 10^{-8}~M_\odot yr^{-1}; L_{acc} \lesssim 10^{-1}~L_\odot$).
At higher rates ($\dot M > 10^{-8}~M_\odot yr^{-1}$), the slope steepens to $\alpha > 1$, 
i.e. a smaller increase in \lha\ per unit increase of \lacc. 
This is because, at very high accretion rates, 
the densities reach a level where the magnetosphere becomes optically thick and the line flux will only emerge from the outer surface.
In addition, increasing outflows may produce \halpha\ absorption.

In the absence of a sufficiently tight correlation to fit a nonlinear trend, we applied a linear regression and find:
\begin{equation}
\rm \log \rm L_{acc} = (1.13 \pm 0.07) \log L_{H\alpha} + (1.93 \pm 0.23),
\end{equation}
where \lha\ and \lacc\ are given in solar units.

We note that there is a significant scatter in the relationship (rms = 0.54),
which is also present for objects where \lha\ and \lacc\ were measured simultaneously.
Indeed correlations of the accretion luminosities with line fluxes have previously been found to
show a large amount of scatter \citep[e.g. see][]{herczeg2008},
which is likely to be caused by a combination of effects 
(e.g. circumstellar absorption, uncertain extinction corrections, line emission not due to accretion).
In \S \ref{sec:errors} we will show that this scatter is the primary source of uncertainty for the accretion rates obtained in this work.

\subsubsection{Mass accretion rate (\macc)}
Having obtained an estimate for \lacc, we now proceed to convert this into an estimate for the mass accretion rate \macc.
We will do this using the same equations as adopted by previous studies in the literature.
In brief, the T Tauri accretion models~\citep[e.g.][]{hartmann1998} assume that the circumstellar disc is truncated at an inner radius \rin\ by the magnetosphere,
and accretion proceeds along magnetic field lines. 
The infalling gas is essentially on a ballistic trajectory, falling on to the star at near free-fall velocities from the distance \rin.
If we assume that the energy released in the impact ``accretion'' shock is reprocessed entirely in the accretion energy \lacc, we may write:
\begin{equation}
\rm L_{\rm acc} \simeq \frac{G M_\star \dot M}{R_\star} (1 - \frac{R_\star}{R_{\rm in}})
\end{equation}
By adopting a standard value $R_{\rm in} = 5\pm2\,R_\star$ \citep{gullbring1998}, we obtain:
\begin{equation}
\label{eqn:macc}
\rm \dot M_{acc} \simeq  1.25\frac{L_{\rm acc} R_\star}{G M_\star}
\end{equation}
Where $R_\star$ and $M_\star$ are taken from the evolutionary tracks in the (\sloanr/\sloanr-\sloani) diagram and \lacc\ is obtained as described above.
The resulting values for \macc\ are listed in Appendix \ref{tab:properties} and plotted in Fig. \ref{fig:hist}.

\section{Archival infrared data}
\label{sec:infrared}

\subsection{2MASS photometry}
Photometry from the near-infrared 2MASS Point Source Catalog \citep{2mass} was retrieved through the Virtual Observatory Conesearch protocol.
We found a match for 152 out of 158 candidates within 0.3\arcsec\ from the IPHAS position
(for 105 objects the match is even better than 0.1\arcsec, which is expected because the IPHAS astrometric calibration is based on the 2MASS catalog).

The remaining 6 candidates are part of close visual binaries. 
The centroid coordinates of these objects differ by more than 1\arcsec\ between both databases,
because the close neighbours were resolved by IPHAS but not by the 2MASS pipeline.
The near-infrared magnitudes for these objects have not been included in our sample,
which is tabulated in Appendix~\ref{tab:candidates}.

\subsection{Spitzer photometry}
We obtained wide-area Spitzer mid-infrared IRAC (3.6, 4.5, 5.8 and 8 $\micron$) and far-infrared MIPS (24 $\micron$) images for the region
from the Spitzer Heritage Archive\footnote{http://sha.ipac.caltech.edu/applications/Spitzer/SHA}.
The IRAC observations were taken in December 2003 (Program ID 37, PI: Giovanni Fazio),
while the MIPS data were obtained in June 2004 (Program ID 58, PI: George Rieke).

The IRAC images cover a box of roughly 35 by 35 arcminutes in the central region 
(shown as a dashed-dotted rectangle in Fig.~\ref{fig:spatial}).
65 of our pre-main sequence candidates (41 per cent) fall inside this box,
62 of which are also covered by the MIPS data (which shows a different footprint).

Basic reduction of the IRAC and MIPS images into artifact-mitigated mosaics 
was performed by a recent version of the Spitzer Science Center Pipeline (versions S18.7 and S18.12).
Photometry was then obtained using the \emph{user-list single frame} option in the MOPEX software (version 18.4.0),
which allows aperture photometry to be performed based on a user-supplied list of target coordinates.

For the IRAC mosaics, we used a 3.6\arcsec\ (6 pixel) aperture radius, and a sky annulus from 14.4 to 24\arcsec\ (24 to 40 pixels).
The MIPS mosaic was reduced using a 13\arcsec\ (5.3 pixel) aperture and a sky annulus from 20 to 32\arcsec\ (8.2 to 13.1 pixels).
The apertures were chosen to avoid flux contamination from nearby sources on one hand, 
while keeping the required aperture corrections as small as possible on the other hand.

Following the IRAC and MIPS Data Handbooks, 
we adopted aperture corrections of 1.112, 1.113, 1.125, 1.218 and 1.170 for the 3.6, 4.5, 5.8, 8.0 and 24 $\micron$ bands, respectively. 
The adopted zeropoints for conversion between flux densities and magnitudes were 280.9, 179.7, 115.0, 64.1 and 7.14 Jy. 

Out of the 65 objects that fell inside the area covered by IRAC, 
all were reliably detected in the four bands at SNR $\gg$ 10.
54 of these objects were also detected in the MIPS 24\micron\ data at SNR $>$ 10
(from the remaining objects, 8 were not reliably detected with MIPS and 3 fell just outside the footprint).

To ensure the reliability of the extracted magnitudes, all sources were inspected by eye.
Four objects were removed from our sample as part of this step: 
three objects because they suffer from severe flux contamination by a very close or bright neighbour 
(IPHAS ID's J213647.63+572954.0; J213657.46+572730.3; J213657.67+572733.1) 
and one object which was partially affected by a gap of missing data in the 4.5\micron\ band (J213816.87+573926.4).

Finally, for those objects for which Spitzer magnitudes were previously reduced by \citet{sicilia2006spitzer},
we compared our magnitudes with the literature values and find an acceptable agreement within $\sim$10 per cent.

The resulting magnitudes for 61 objects are tabulated in Appendix~\ref{tab:spitzer} 
and plotted as colour-colour diagrams in Fig.~\ref{fig:ir}.
We repeated the same reduction steps for the list of WTTS objects from SA06
and included them in Fig.~\ref{fig:ir} (black crosses) for comparison.

\section{Results}
\begin{figure*}
\centering
\includegraphics[width=\textwidth]{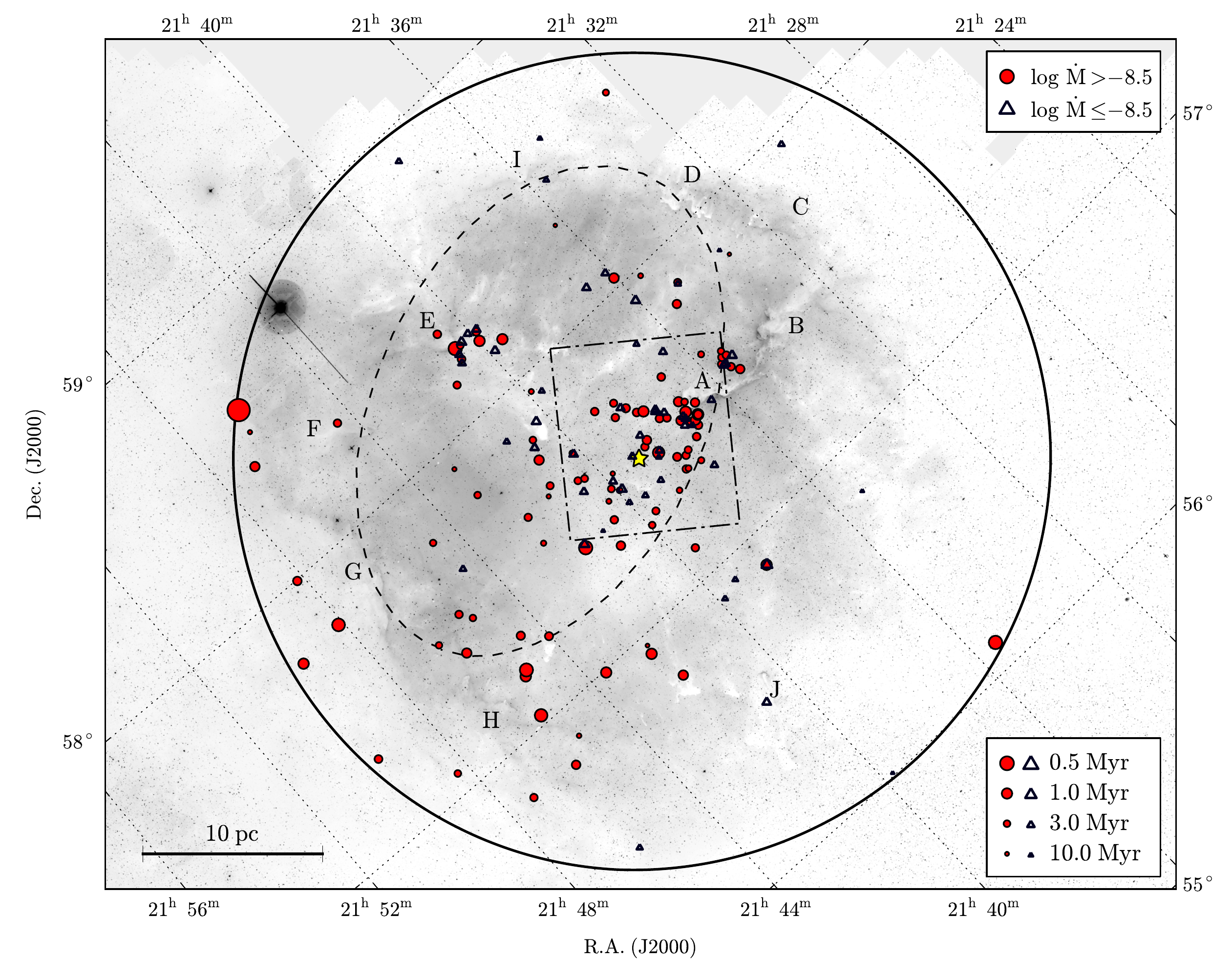}
\caption{\label{fig:spatial} Spatial distribution of pre-main sequence candidates (circles and triangles).
The size of the symbol denotes the age, the shape/colour denotes the mass accretion rate.
The large solid circle shows the 3\deg\ diameter circular region considered in our work, centred on the O6.5V-type star HD\,206267 (yellow star symbol).
The background shows an inverted greyscale \halpha\ mosaic of the region.
The prominent bright-rimmed clouds in the region are indicated with letters A-J as assigned in \protect\citet{pottasch1956} and \protect\citet{weikard1996}.
The expanding ring of molecular material on which the clouds are thought to be located \protect\citep{patel1995} is indicated by the dashed ellipse.
The central region covered by Spitzer data is indicated by the dashed-dotted rectangle.
We note that a large number of candidates are clustered in front of the bright-rimmed clouds A and E.
} 
\centering
\includegraphics[width=0.92\textwidth]{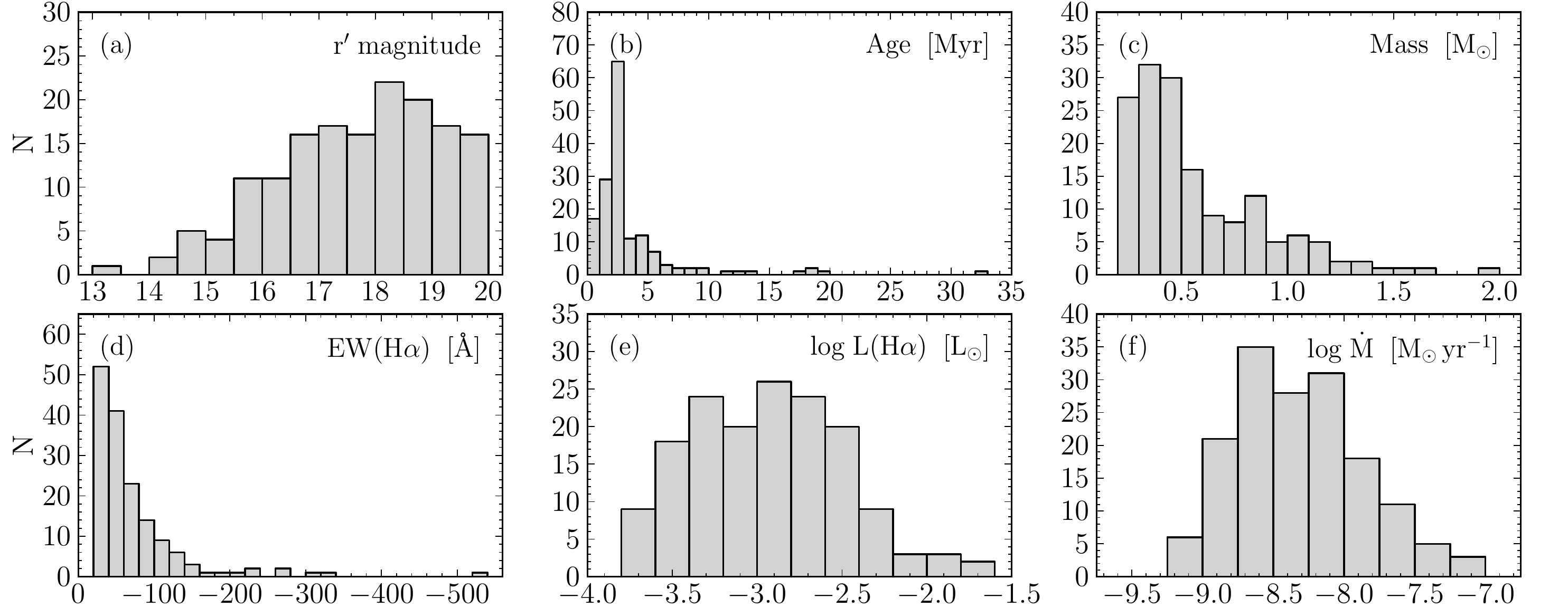}
\caption{\label{fig:hist} Distribution of stellar and accretion parameters as tabulated in Appendix A-B.} 
\end{figure*}
\begin{figure*}
\includegraphics[width=\linewidth]{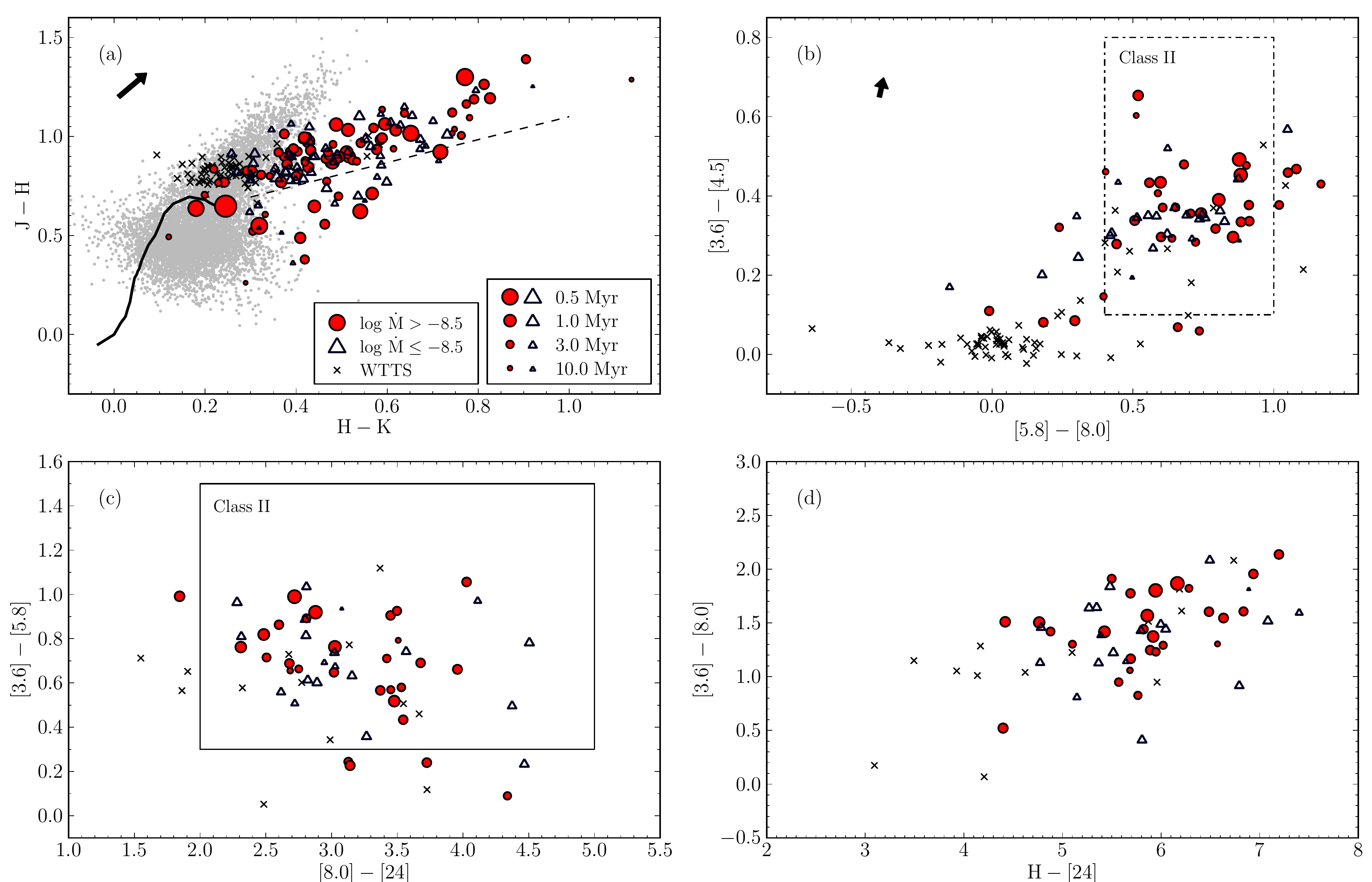}
\caption{\label{fig:ir} 
Infrared colour-colour diagrams for the pre-main sequence candidates (red filled circles and blue open triangles) 
and weak-lined cluster members with \ewha\ weaker than -10\,\AA\ from \protect\citeauthor{sicilia2006} (crosses).
(a) Near-infrared colours for stars with ``AAA''-quality data in the 2MASS database. Grey dots show a sample of 8\,000 field stars in the region.
The dashed line shows the typical position of unreddened T Tauri stars determined by \protect\citet{meyer1997},
while the solid line denotes the main sequence for dwarfs due to~\protect\citet{bessellbrett}.
The arrow shows the reddening vector for $\rm A_V=1$.
(b) Spitzer IRAC colours for candidates in the central region (see text).
The dashed-dotted box shows the typical colours for T-Tauri stars as predicted by the models of \protect\citet{dalessio2005} for $\rm log\,\dot M = -8\ M_\odot\,yr^{-1}$.
The arrow shows the reddening vector for $\rm A_V=3$.
(c) Spitzer IRAC colours combined with the MIPS [24] band. 
The solid box shows the typical colours for Class II objects found by \protect\citet{muzerolle2004}.
The objects which fall below the box are candidate Transition Objects.
Note that most of the WTTS objects were not detected in the [24] band and are therefore not included in the diagram.
(d) Combination of 2MASS, IRAC and MIPS colours.
All diagrams show positive but statistically insignificant correlations between the colours, accretion rates and ages.
}
\end{figure*}

	The spatial distribution of our sample is shown on top of the \halpha\ mosaic in Fig.~\ref{fig:spatial}.
The size and shape of the symbols reflect the estimated accretion rates and ages.
We note that most candidates are clustered in front of the bright-rimmed molecular clouds A, B and E \citep[nomenclature from][]{pottasch1956}.
In the final section of this paper we will argue that the location and dispersion of these clusters suggest that their formation was triggered by the central O-type star.

	Histograms of the derived parameters are shown in Fig.~\ref{fig:hist}.
We note that the age distribution (panel b) shows a median age of 2.3~Myr.
This is in good agreement with the median age of 2.6~Myr for the confirmed classical T Tauri members from SA06.
It is also consistent with the expansion age of the gas derived from CO radial velocity measurements \citep[2-3~Myr,][]{patel1995},
but possibly younger than the upper main sequence turnoff age \citep[$\sim$4-7~Myr][]{marschall1990,contreras}.

	The mass distribution (panel c) shows the expected power-law increase towards lower masses.
When compared with a Kroupa IMF, our sample shows a deficit of candidates lower than \mass\ $<$ 0.3\,\msol\ and beyond \mass\ $>$ 1.5\,\msol,
which can be explained by the imposed magnitude limits.

	The equivalent widths (panel d) range from -22 to -525~\AA,
while the mass accretion rates (panel f) range from $\rm 10^{-9.2}$ to $\rm 10^{-7.0}~M_{\odot} yr^{-1}$.
These ranges are consistent with values for strong accretors in other regions.

	Infrared colour-colour diagrams are shown in Fig.~\ref{fig:ir}.
In the near-infrared 2MASS diagram, we note that many candidates are well separated from the general locus of stars due to an H-K excess 
(91 per cent showing H-K $>$ 0.3; 39 per cent showing H-K $>$ 0.5).
The emission is commonly thought to be caused by emission from the inner dust rim of the accretion disc.
Some objects are not separated from the main locus,
which might give the impression that they do not have a disc.
However, the accretion disc models due to \citet{meyer1997} show that a near-infrared excess 
may disappear in the case of a large inner-hole size or an edge-on inclination angle,
therefore the absence of an H-K excess does not necesarilly imply the absence of a disc.
(In fact we did not include 2MASS colours as part of our selection criteria for this reason.)

	Indeed, the objects for which mid- and far-infrared Spitzer data is available 
all show colour excesses consistent with having accretion discs
(panels b-c-d in Fig.~\ref{fig:ir}). 
Five objects are seen to show strong infrared excess only at 8 and 24\micron\ but not at shorter wavelengths,
indicating that dust is absent from the inner regions of the disk (i.e., candidate ``transition objects'';  
J213709.09+572548.6, J213825.90+574204.8, J213830.30+573255.2, J213955.69+571638.2 and J214035.33+572309.6).

\section{Validation}
\label{sec:validation}

\subsection{Are our candidates genuine T-Tauri stars?}
Several arguments allow us to assume that the vast majority of our candidates are genuine T-Tauri members of IC\,1396:
\begin{enumerate}
 \item all objects fall within the 0.1-100~Myr isochrones placed at the distance and extinction of the region;
 \item the stellar ages of 2-3~Myr agree with literature findings (SA06);
 \item the 2MASS near-infrared colours are consistent with the T Tauri locus by \citet{meyer1997};
 \item the Spitzer mid- and far-infrared colours are consistent with accretion disc models by \citet{dalessio2005} and observations by \citet{muzerolle2004};
 \item the candidates are strongly clustered;
 \item 49 of the objects are known in SIMBAD, all having types consistent with young stars (T~Tau, IR or X-ray).
\end{enumerate}

Moreover, as part of the ongoing IPHAS spectroscopic follow-up campaign, 
five of our candidates with very strong emission-lines 
have already been observed by Valdivielso, Bouy \& Martin (submitted; objects J213528.42+575823.1, J213545.88+573640.1, J213938.83+575451.4, J214547.74+564845.8 and J214625.99+572828.9).
These objects were not previously studied by Sicilia-Aguilar et al.
All five spectra show M-type SEDs with strong \halpha\ emission (ranging between -105 and -360\,\AA, in agreement with our estimates) 
as well as emission from the Ca {\sc II} infrared triplet, which is typical for classical T Tauri type stars.

\subsection{Contamination rate and completeness}

\begin{figure}
\includegraphics[width=\linewidth]{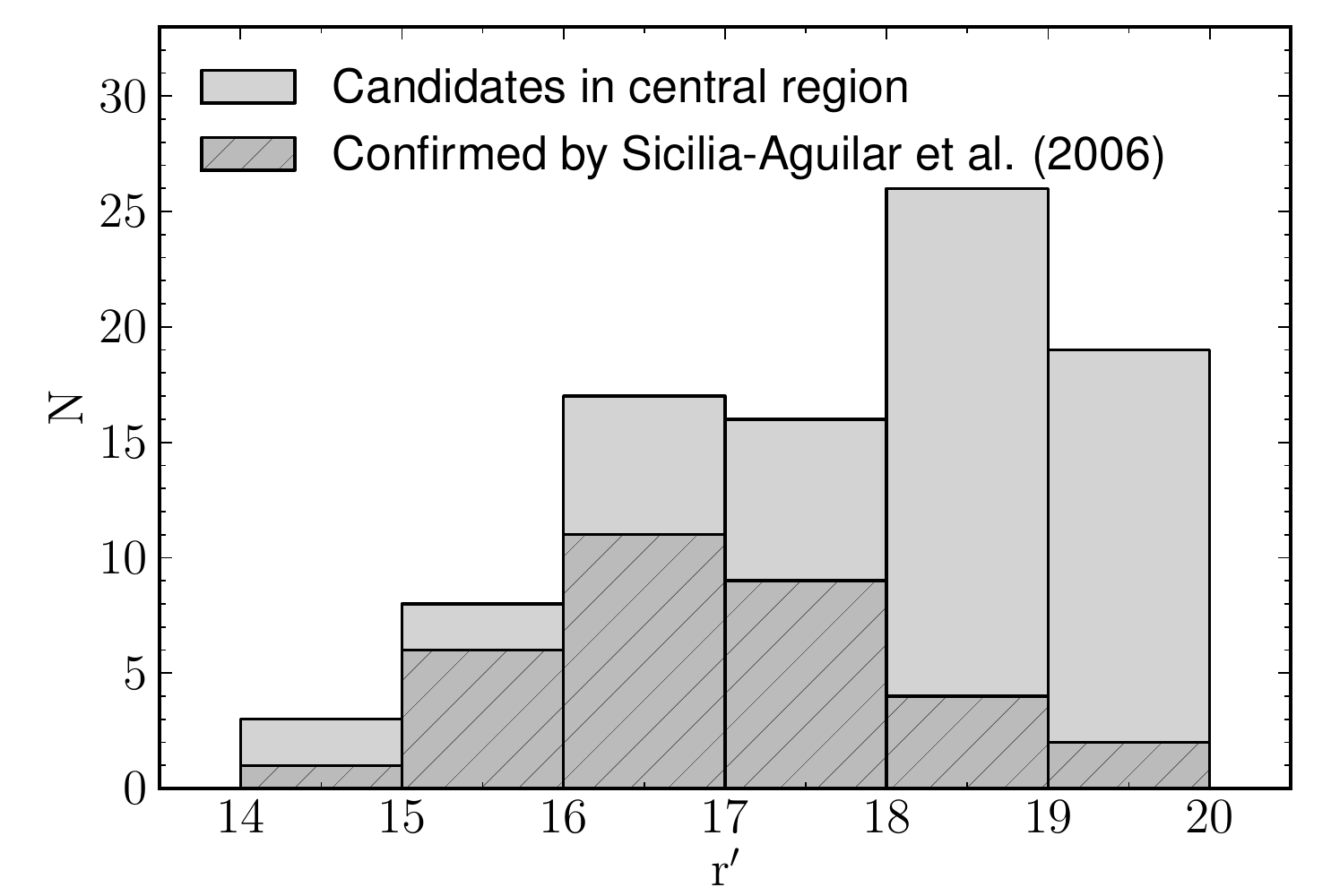}
\caption{\label{fig:aguilar_confirm} 
Spectroscopic confirmation of T-Tauri candidates as a function of magnitude.
Only objects in the centre of IC\,1396, defined as the region within 0.5\deg\ from the ionizing star, are included in this plot.
This corresponds to the region which was surveyed by SA06.
The majority of objects brighter than $\mathrm{r'} < 18$ are spectroscopically confirmed T-Tauri stars.
In addition, our work finds many new low-mass candidates at fainter magnitudes due to the greater depth of the IPHAS survey.
}
\includegraphics[width=\linewidth]{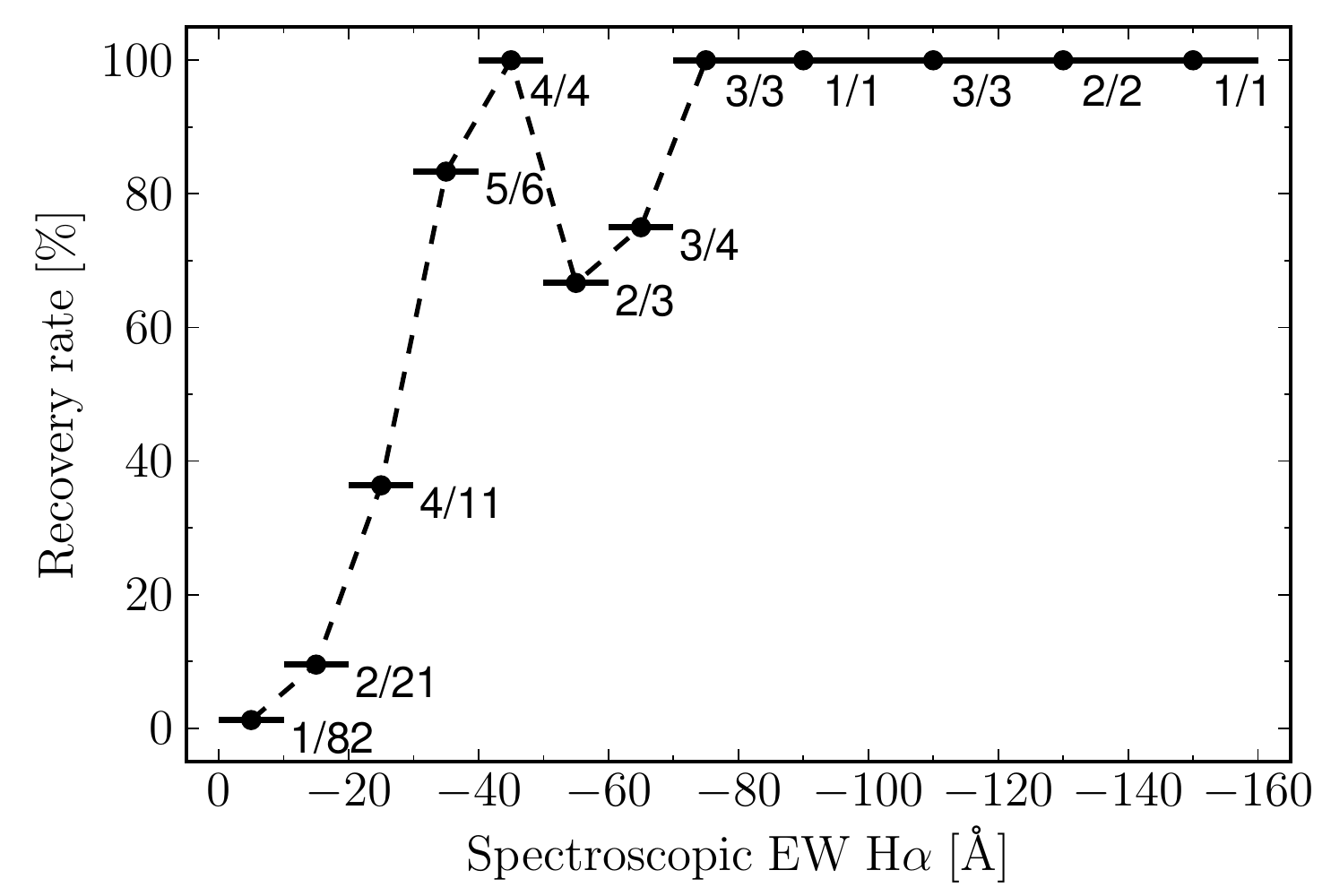}
\caption{\label{fig:aguilar_ew} Fraction of members from \protect\citet{sicilia2006} which are also recovered in our work,
given as a function of the spectroscopic equivalent width.
We included literature objects which are listed as ``confirmed'' or ``probable'' members
and for which a low-resolution spectroscopic equivalent width is available (141 objects).
We find that members with \halpha\ lines weaker than $-30$\,\AA\ are not recovered because they fall below the selection threshold that we used.
}
\end{figure}

While we can assume that the vast majority of candidates are genuine members, 
it is likely that our method includes at least some contamination from foreground or background emission-line objects. 
We may constrain this contamination rate as follows.

The surface density of our sample is equal to 22.3 candidates/\sqdeg\ (158 candidates in 7.1~\sqdeg). 
The density is not uniform in the entire area.
On one hand, the region within 1\deg\ from HD\,206267 contains 135 candidates and has a density of 43.0 objects/\sqdeg.
On the other hand, the region just south of IC\,1396 shows a density of only $\sim$3.2 objects/\sqdeg.
If we adopt the latter number as an upper limit for the rate of contamination, then our sample would contain at most 23 non-members (15 per cent),
which is similar to the number of objects that fall below the T Tauri locus in the 2MASS diagram.

In reality, it is very likely that emission-line stars in the outskirts are genuine members of the surrounding 
Cepheus OB2 association and that the contamination rate is lower. 
We conclude that the sample is sufficiently clean to allow it to be used for initial statistical investigations.

A further test to validate our sample is to compare our objects with the members confirmed by SA06.
In Fig.~\ref{fig:aguilar_confirm} we plot the total number of IPHAS candidates in the central region, defined as 
$\sim$0.5\deg\ from HD\,206267, and indicate the number of these objects which have been found previously.
We find that 18 out of 28 of our candidates brighter than \sloanr\ $<$ 17 have already been found and confirmed.
At fainter magnitudes, we find many new candidates.
This can be explained from the fact that the photometric selection of candidates for spectroscopic follow-up by Sicilia-Aguilar et al.
concentrated on stars with V=15-19 (R$\approx$14-18), while we considered objects up to \sloanr = 20.

To estimate the completeness of our sample, 
Fig.~\ref{fig:aguilar_ew} shows the fraction of objects from SA06 which were successfully retrieved by our work. 
We find that for objects with \ewha\ stronger than -30\,\AA, 24 out of 27 objects were successfully recovered (89 per cent).
From the remaining 3 objects, 
one is heavily affected by source confusion and was rejected by our criteria because the pipeline classified it as ``non-stellar'' (SA06 ID 13-1877). 
The two other objects (14-1017 and 92-393) fall just below the selection threshold, 
which can be explained by a small decrease in the line intensity due to natural variability.

We conclude that, compared to the extensive spectroscopic survey, our method appears to be equally complete for objects with lines stronger than -30\,\AA.
This is consistent with the fact that the selection threshold lies near the -25\,\AA\ boundary at the cluster reddening (compare Fig.~\ref{fig:ccd_aguilar} and \ref{fig:ccd_selection}).

In brief, our comparison with the literature indicates that we obtained a reliable and nearly equally complete picture of strong accretors in the region. 
More precise detectability limits as a function of mass and age will be given in \S \ref{sec:discussion}.

\section{Uncertainties in the derived parameters}
\label{sec:errors}
Our stellar parameters suffer from uncertainties 
caused by errors in the input data and the physical assumptions.
In what follows we will discuss two types:
\begin{enumerate}
 \item {\em non-systematic uncertainties}: the fact that individual stars may have a true extinction or distance 
which deviates from the assumed mean values,
thereby introducing errors in the derived parameters relative to the other objects in the sample;
 \item {\em systematic uncertainties}: the fact that the assumed mean distance/extinction values or the evolutionary tracks may be shifted in a way 
 that affects the derived parameters for all stars in a systematic direction.  
\end{enumerate}
In what follows we will also discuss the effect of unresolved binaries 
and investigate the age spread.

\subsection{Non-systematic uncertainties}
\label{sec:montecarlo}

\begin{table*}
\caption{Summary of the non-systematic uncertainties.
The mean standard deviations ($\bar\sigma$) are computed from the individual $1\sigma$-errors given for each object in Table~\ref{tab:properties}.
}
\begin{tabular}{llccccccc}
\hline
Source & Simulated & $\rm \bar\sigma(Age)$  &  $\rm \bar\sigma(Mass)$ & $\rm \bar\sigma(Radius)$ & $\rm \bar\sigma( EW_{H\alpha} )$ & $\rm \bar\sigma( \log L_{H\alpha} )$ & $\rm \bar\sigma( \log \dot M )$ \\
& uncertainty & [Myr] & [M$_\odot$] & [R$_\odot$] & [\AA] & [$\rm L_\odot$]	&  [$\rm M_\odot\,yr^{-1}$]  \\
\hline
Photometry & $\rm \bar\sigma(r') = 0.01$	& 0.18 & 0.01 & 0.04 & \bf 4.16 & 0.02 & 0.04 \\
Extinction & $\rm \sigma(A_v) = 0.55$	& \bf 0.78 & \bf 0.10 & \bf 0.17 & 3.89 & \bf 0.19 & 0.20 \\
Distance & $\rm \sigma(d) = 20~pc$	& 0.20 & 0.01 & 0.05 & 0.00 & 0.02 & 0.04 \\
\lha-\lacc\ relation & $\rm \sigma(\log L_{acc}) = 0.54$	& - & - & - & - & - & \bf 0.54 \\ 
\hline
Combined & All of the above   			&	0.89 & 0.10 & 0.18 & 5.97 & 0.20 & 0.59 \\
\hline
\label{tab:errors}
\end{tabular}

\caption{Summary of the simulated systematic uncertainties.
Each column shows how shifting one of the input assumptions affects the mean values of the stellar parameters in our sample. 
}
\begin{tabular}{llccccccc}
\hline
Source & Simulated & $\rm \mu(Age)$  &  $\rm \mu(Mass)$ & $\rm \mu(Radius)$ & $\rm  \mu(EW_{H\alpha})$ & $\rm \mu(\log L_{H\alpha})$ & $\rm  \mu(\log \dot M) $ \\
& shift & [Myr] & [M$_\odot$] & [R$_\odot$] & [\AA] & [$\rm L_\odot$]	&  [$\rm M_\odot\,yr^{-1}$]  \\
\hline
Extinction shift & $\rm \bar{A}_V\ +\ 0.5$ & $-0.00$ & +0.10 & +0.11 & $-3.71$ & +0.18 & +0.17 \\
" 				 & $\rm \bar{A}_V\ -\ 0.5$ & +0.20 & $-0.08$ & $-0.08$ & +3.33 & $-0.18$ & $-0.16$ \\

Distance shift & $\bar{d}\ +\ $100~pc & $-0.71$ & +0.00 & +0.17 & - & +0.09 & +0.15 \\
" 			   & $\bar{d}\ -\ $100~pc & +1.10 & $-0.01$ & $-0.14$ & - & $-0.11$ & $-0.15$ \\
Model tracks shift & $\rm r' + 1.0;\ i' + 0.5$ & $-1.98$ & $-0.26$ & +0.19 & - & - & +0.27 \\
\ " & $\rm r' - 1.0;\ i' - 0.5$ & +4.43 & +0.29 & $-0.07$ & - & - & $-0.25$ \\
\hline
\label{tab:errors2}
\end{tabular}

\end{table*}

Constraining the uncertainties using analytical error propagation
would be complex and error-prone. 
Instead we opt to investigate the uncertainties using a Monte Carlo approach which works as follows.

For each star in our sample, we generated 10\,000 sets of artificial ``clones''.
Each clone is perturbed by randomly adding Gaussian noise to the input photometry and 
to the physical assumptions (distance, extinction, \lha-\lacc\ regression).
For the photometry, the noise is added according to the Poissonian errors listed in Table~\ref{tab:candidates}. 
For the error in the extinction, we assume the literature value $\rm \sigma(A_v) = 0.55$ (SA05). 
The adopted error in distance is the approximate diameter of the region: $\rm \sigma(d) = 20~pc$.
Finally, the error in the \lha-\lacc\ regression is the rms scatter we found previously: $\rm \sigma(L_{acc}) = 0.54$.

After generating a set of clones, we then recompute the stellar parameters for each clone; 
resulting in a set of 10\,000 estimates per parameter per star.
These sets can be considered to be samplings of probability distributions.
We computed the mean and 1-sigma standard deviation\footnote{In the case of the stellar age, where the probability distribution is heavily skewed, 
we list the median age along with the 25\% and 75\% quartiles.
The logarithm of the age does have a symmetric uncertainty distribution, but we choose to print the linear age for readability.}
from each of these sets,
which are the numbers given for each object in Table~\ref{tab:properties}.

We summarized the uncertainties in Table~\ref{tab:errors} by showing the mean 1-sigma uncertainty for the entire sample per output parameter and per source of error.
The summary shows that the determination of the stellar parameters suffers mostly from the uncertainty in the extinction.
However, the final uncertainty in the accretion rate is dominated by the scatter in the \lha-\lacc\ regression.
Because individual uncertainties add quadratically to form the combined uncertainty (i.e., $\rm \sigma_{A+B} = \sqrt{\sigma^2_A + \sigma^2_B}$ if A and B are independent),
a single largest source of uncertainty tends to dominate the total uncertainty,
therefore a better estimate of the extinction would not significantly improve the accuracy of the accretion rates. 

We conclude that the accretion rates are constrained within an uncertainty of $\sim$0.6 dex relative to the other stars in the sample,
mainly due to the scatter in the relationship between \lha\ and \lacc.
The ages are constrained within $\sim$1~Myr and masses within $\sim$0.1\,\msol.

\subsection{Systematic uncertainties}
We investigated the effect of systematic uncertainties
by introducing shifts in:
\begin{enumerate}
 \item the assumed mean extinction ($A_V \pm 0.5$);
 \item the assumed mean distance ($d \pm 100$ pc);
 \item the position of the evolutionary tracks.
\end{enumerate}
We recomputed the stellar parameters with each of these shifts introduced,
and summarized the results in Table~\ref{tab:errors2}.
We find that a shift in the assumed extinction shifts the masses by $\pm 0.1$\,\msol,
while a shift in the distance shifts the ages by $\pm 1$~Myr.

For simulating errors in the model track positions, we adopted the conclusions by \citet{lawson2001} 
who found that the position of mass tracks and isochrones in colour-magnitude diagrams differ by up to $\sim$1.0 mag and $\sim$0.5 colour between different models. 
Our simulations show that shifting the \citet{siess} tracks downwards by (\sloanr,\sloanr-\sloani)$+$(1.0,0.5) makes the ages on average 2~Myr younger and masses 0.3~\msol\ lighter. 
Shifting the tracks upwards by (\sloanr,\sloanr-\sloani)$-$(1.0,0.5) yields ages which are 4~Myr older and masses 0.3~\msol\ higher.
The ages appear to be very uncertain for this reason, and may perhaps be wrong by a factor two in an absolute sense \citep[see also][]{naylor2009,baraffe2009}.

\subsection{Unresolved binaries}
We cannot exclude that a fraction of our candidates are very close unresolved binaries (e.g. having a separation of less than $\sim$1\arcsec, i.e. $\sim$1000~AU).
We investigated this effect by simulating what happens when the luminosity of a star is over-estimated by a factor two while the colour remains unchanged 
(the case of an unresolved equal-mass binary system).
We find that such case leads to an age estimate which is on average 4~Myr too young and mass accretion rates which are 0.5 dex too low.

We did not include this effect in the simulations because the fraction of unresolved binaries is highly uncertain.
Observations by \citet{bouwman2006} suggest that close binaries lose their disk significantly faster than single stars, 
which means that the number of binaries in our sample could be very low.
If this is not the case, unresolved binaries may be the single most important source of error in our work.

\subsection{Is the spread in ages real?}

\begin{figure}
\includegraphics[width=\linewidth]{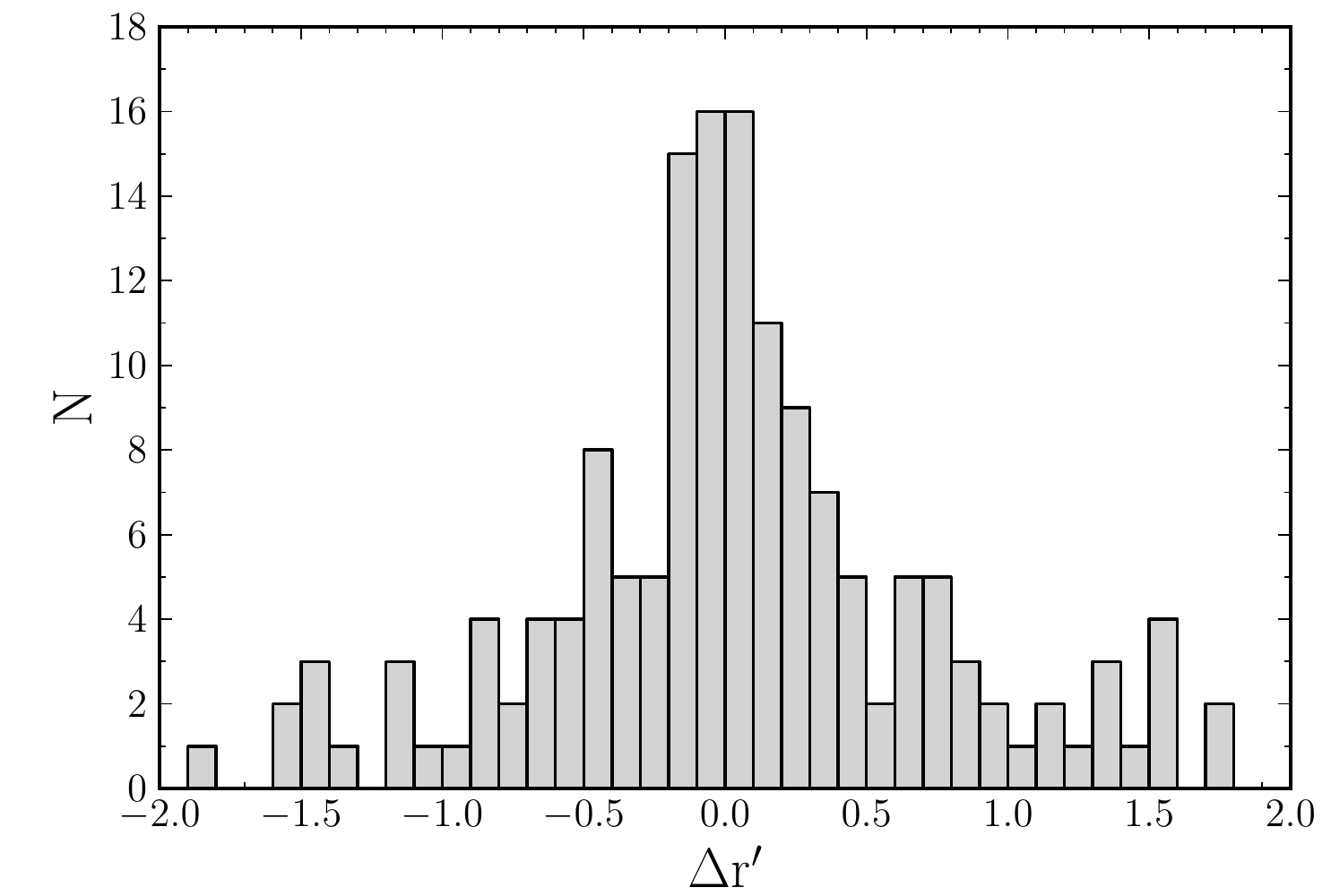}
\caption{\label{fig:agespread} Spread in $r'$ magnitude around the 2.5~Myr isochrone shown in Fig.~\ref{fig:tracks}.
It is unclear whether the scatter can entirely be explained by photometric variability and uncertainties, 
or if a genuine age spread exists in the region.}
\end{figure}

The estimated ages of our candidates range between 0.2 and 32~Myr, with a median of 2.3 and standard deviation 4.0.
The spread is significantly larger than the standard deviation of $\sim$0.9 which is predicted to occur 
due to the combined uncertainties in the photometry, extinction and distance (Table~\ref{tab:errors}).
Does this mean that a true age spread exists in the region?

We must take into account that T Tauri stars have been observed to change in brightness by as much as one magnitude on timescales of only days \citep[e.g.][]{eiroa2002},
which is thought to be explained by stellar spots, variable accretion, or variable obscuration by circumstellar dust.
In the colour-magnitude diagram presented earlier (Fig.~\ref{fig:tracks}), 
we found that the objects are widely scattered around the 2.5~Myr isochrone, 
with 79 per cent deviating by $\Delta r'\,\geq\,0.1$ from the isochrone 
and 37 per cent showing $\Delta r'\,\geq\,0.5$ (Fig.~\ref{fig:agespread}).  
This is slightly larger than the variability of stars in this region recently reported by \citet{sicilia2010} based on multi-epoch V band magnitudes,
where the authors found 65 per cent showing a variability $\geq 0.1$ mag and 16 per cent $\geq 0.5$ mag.
This indicated that the observed variability may explain a significant part of the age spread, although perhaps not entirely.

If the effect of unresolved binaries play a significant role, 
it is likely that the combined effect of uncertainties and variability does explain the entire age spread.
However, it will always be difficult to exclude that perhaps a small true age spread 
-- heavily masked by the noise -- is hiding in the large apparent spread.

\section{Discussion}
\label{sec:discussion}
In this section we provide a discussion on the implications of our results.
First we investigate how our mass accretion rates depend on age and mass, 
then we discuss the implications for the history of IC\,1396 and the role of the massive star HD~206267.

\subsection{Accretion rates as a function of mass and age}

\subsubsection{The \macc-\mass\ relationship}

\begin{figure}
\includegraphics[width=\linewidth]{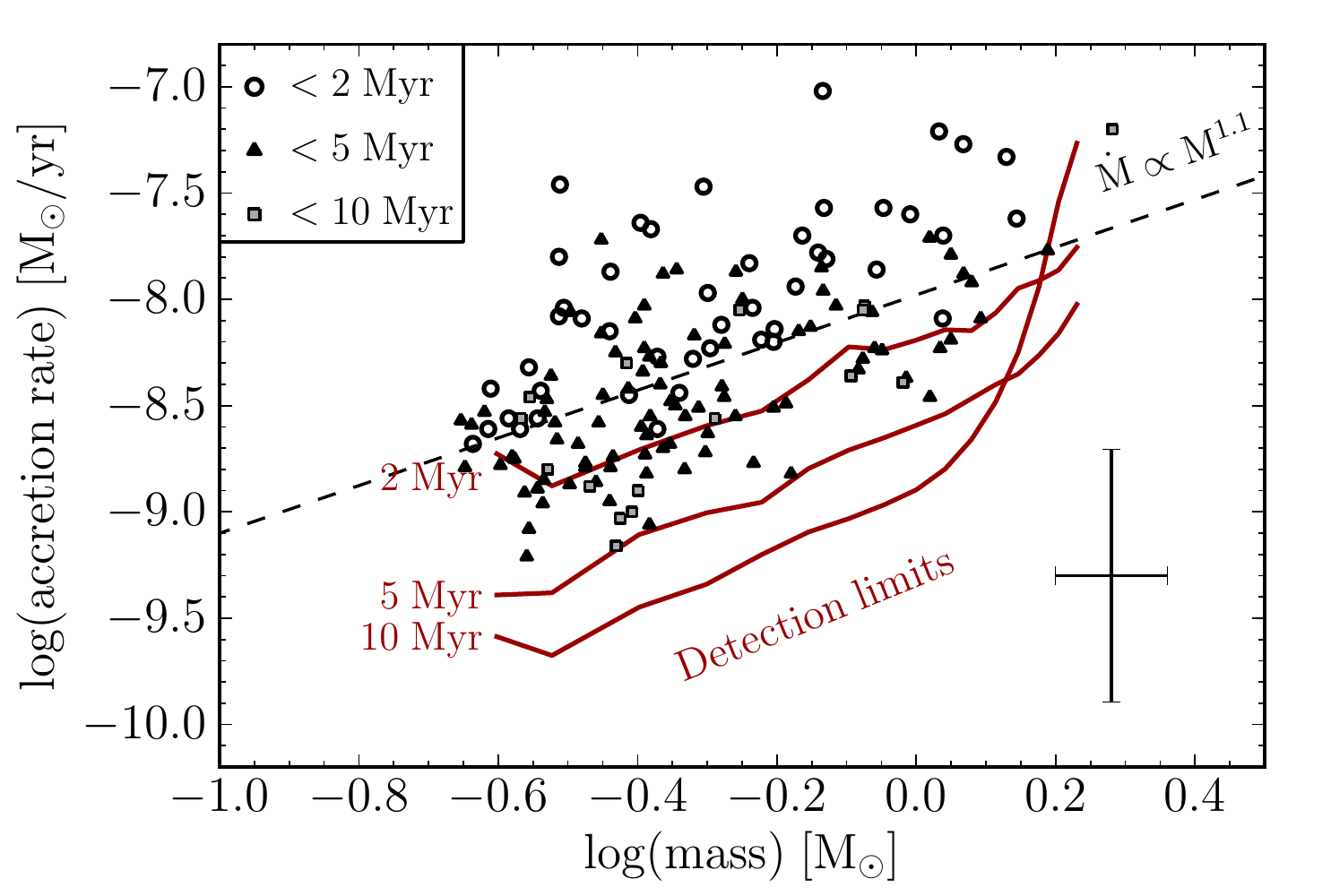}
\caption{\label{fig:massdep} Distribution of mass accretion rates as a function of stellar mass for our T-Tauri candidates.
We show candidates younger than 2~Myr (open circles), 5~Myr (filled triangles) and 10~Myr (filled squares). 
The solid red lines show the detectability limits as a function of age and mass.
The dashed line shows a linear regression through the data, which is heavily affected by the detectability limits.}
\end{figure}

Fig.~\ref{fig:massdep} shows the distribution of mass accretion rates as a function of stellar mass. 
We fitted the $\rm \dot M_{acc} \propto M_\star^\alpha$ power-law relationship and find a slope $\alpha = 1.1\pm0.2$ (all objects)
or $\alpha = 1.3\pm0.2$ (objects younger than 2~Myr), although with a large spread in \macc\ for any \mass. 

The slope is less steep than recently claimed in literature, 
where different authors found the accretion rate to correlate roughly with the square of the stellar mass; 
$\alpha = 1.8-2.3$ \citep{muzerolle2003,mohanty2005,natta2006,herczeg2008} and recently as steep as $\alpha = 3.1$ \citep{fang2009}.

The apparent inconsistency with the literature may be explained by two reasons.
First, our dataset does not include very low mass stars or brown dwarfs,
which form a significant part of the mass range considered in several of the previous studies.
We do not rule out that our slope would have been steeper if we had we been able to detect such objects at the distance of IC\,1396.
Likewise, we note that our data covers only $\sim$2 magnitudes in mass accretion rates.

Second, following \citet{clarke2006} and \citet{mayne2010},
the determination of the slope is affected by method-dependent systematic effects,
namely:
\begin{enumerate}
\item detectability limits in the (\macc,\mass) plane;
\item intrinsic correlations between \mass/\radius/\lum.
\end{enumerate}

An advantage of our photometric method is that it is straightforward to derive its detectability limits,
because all objects were selected in a homogeneous way from a precisely defined threshold in the (\sloanr-\sloani/\sloanr-\halpha) plane.
The detectability limits derived from this threshold, computed for objects of different ages,
are shown as red solid lines in Fig.~\ref{fig:massdep}.
We find that the limits correlate with the mass in a similar way as the accretion rates ($\alpha \simeq 1.2$);
i.e. it is easier to detect smaller accretion luminosities around lower mass (less luminous) stars.

It is worth noting that the compilation by \citet{natta2006}, 
which shows a steeper dependency ($\alpha = 1.8 \pm 0.2$),
is also affected by a steeper slope in the detection limit \citep[$\alpha \simeq 2$, see][]{clarke2006}. 
The selection effects of the other studies, however, are less clear.

The second effect suggested by \citet{clarke2006} 
are the tight natural correlations between \mass/\radius/\lum\ for stars of a similar age.
Because we derive all stellar parameters using a single method (namely, model tracks in the IPHAS diagrams),
the uncertainties of these parameters are not independent.
Using the Monte Carlo simulations discussed previously, 
we were able to determine the covariance between the mass and the accretion rate for each star (using the set of 10\,000 clones disturbed with errors). 
We find strong positive covariances, 
i.e. the uncertainty in the accretion rate tends to be correlated 
and have the same sign as the uncertainty in the stellar mass.
The mean covariance-to-variance ratio is:
\begin{equation}
\langle \rm {{covariance(\log \dot M_{acc}, \log M_\star)}\over{variance(\log M_\star)}} \rangle = 1.7.
\end{equation}
This ratio can be interpreted as follows: in a sample of (equal-mass) stars where the apparent mass range would be entirely due to uncertainty, 
we would expect to retrieve a default slope $\alpha=1.7$ driven purely by the effect of correlated uncertainties.
Fortunately, the mass range considered in this paper is several times larger than the estimated mass uncertainty, 
therefore the impact of the effect is limited. 
Our simulations suggest that the effect accounts for an overestimation of the slope by $\sim$10 per cent.

We conclude that we do not find evidence to suggest that accretion rates correlate with the square of the stellar mass in the range 0.2-2.0\,\mass.
We stress, however, that we do not rule out such relationship from this dataset. 
We do emphasize the importance of including detection limits and covariance estimates in the discussion.

\subsubsection{The \macc -Age relationship}
\begin{figure}
\includegraphics[width=\linewidth]{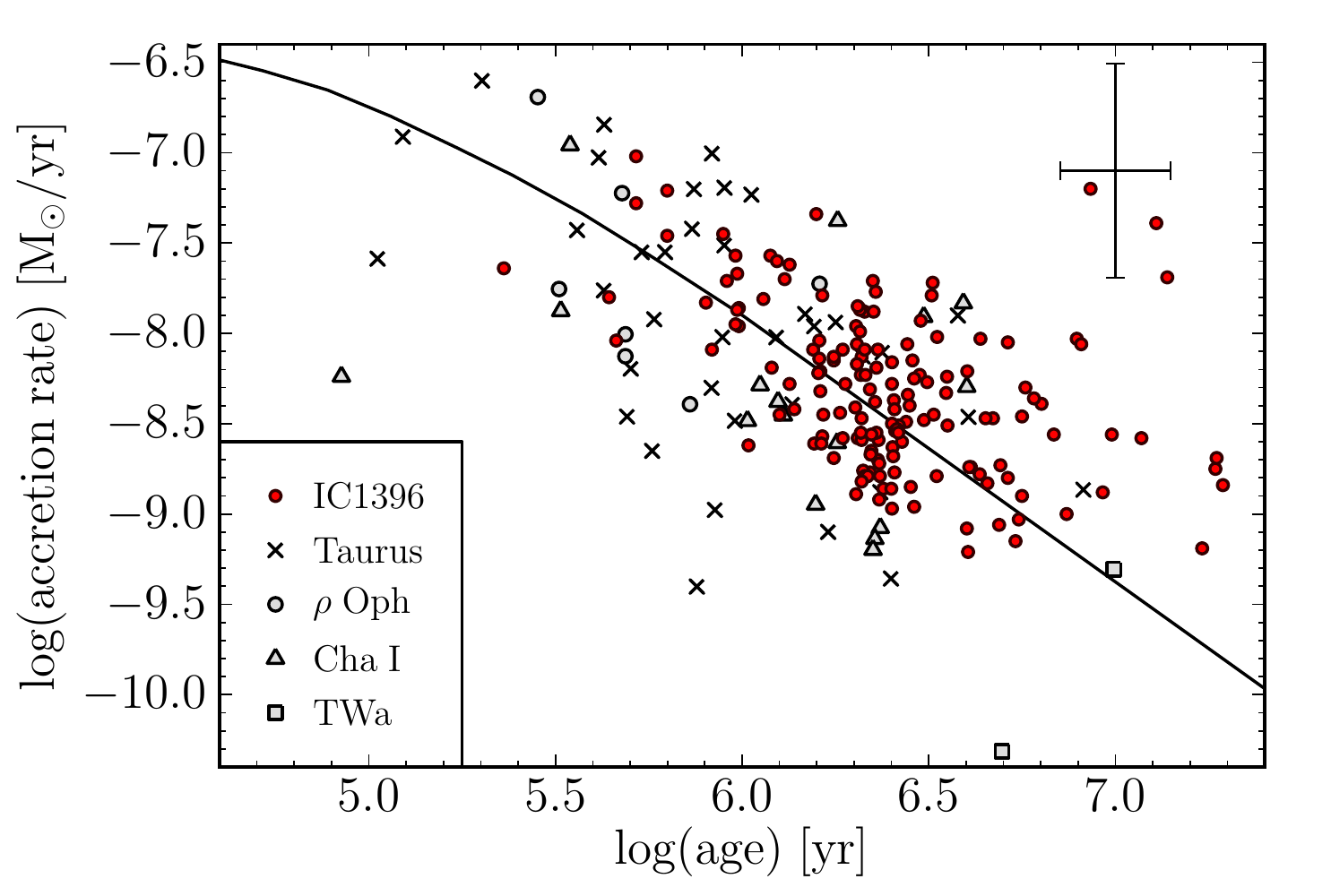}
\caption{\label{fig:agedep} Distribution of mass accretion rates as a function of stellar age for our T-Tauri candidates (red filled circles),
compared with literature values collected by~\protect\citet{muzerolle2000}.
The solid line indicates the evolution for models of viscous disc evolution due to~\protect\citet{hartmann1998}.
The average 1-sigma error is shown in the top right.}
\end{figure}

Figure~\ref{fig:agedep} shows the distribution of accretion rates as a function of stellar age,
together with spectroscopic results for other regions taken from literature.
We find $\rm\dot M_{acc}\propto\rm Age^{-0.9\pm0.1}$ for our sample,
which is in excellent agreement with the trend for literature data: 
$\dot M_{acc}~\propto\rm Age^{-1.0\pm0.2}$ \protect\citep{muzerolle2000},
although slightly less steep than expected theoretically \citep{hartmann1998}.

However, the \macc$-$Age relationship is subject to the same statistical effects discussed in the previous section.
First, in Fig.~\ref{fig:massdep} we already showed that lower mass accretion rates are more easily detected around older (less luminous) stars.
Second, intrinsic correlations in the parameter space yield a negative covariance-to-variance ratio:
\begin{equation}
\langle \rm {{covariance(\log \dot M_{acc}, \log Age)}\over{variance(\log Age)}} \rangle = -0.6.
\end{equation}
i.e. in a sample of coeval stars where the age spread is dominated by errors, 
we expect to find a default relationship $\dot M_{acc}\propto\rm Age^{-0.6}$.

The effect can easily be understood from the fact that the age and the radius 
are both estimated from the same colour-magnitude diagram.
For example, when the magnitude of an object is over-estimated (e.g. due to an error in the photometry or distance),
the object will be located higher up in the colour-magnitude diagram near a younger isochrone,
which will lead to an under-estimated age and at the same time an over-estimated radius 
(because $\rm \log\,R_\star$ is roughly proportional to $\rm \log\,Age$ in the model tracks). 
In turn, the larger radius will automatically lead to an over-estimated accretion rate (cf. Equation~\ref{eqn:macc}).

We have previously shown that, unlike masses, 
the model ages are highly uncertain and the age spread is likely to be explained, 
at least for a significant part,
by uncertainties and variability.
Therefore the effect of correlated uncertainties is likely to play a very significant role.
The same effect is likely to occur in spectroscopy-based studies as well.

We conclude that the \macc$-$Age relationship for a single region must be interpreted with great care, 
especially where the age spread is uncertain. 
Large samples of accretion rates accross regions, 
spanning a large range of truely different ages, 
are essential before conclusions may be drawn on the relationship.

\subsection{Role of the ionizing star HD\,206267 (O6.5V)}

\subsubsection{The influence of massive stars}
It is often claimed that radiation and stellar winds from O-type stars 
may have a profound effect on nearby star formation.
On one hand, a massive star may disrupt star formation 
by dispersing the local molecular cloud
and removing circumstellar material from newly formed stars \citep[\emph{UV photoevaporation}, e.g.][]{johnstone1998}.
On the other hand, expanding ionization fronts may act to compress the gas around 
the periphery of an H{\sc ii} region which then collapses to form stars 
\citep[\emph{collect and collapse},][]{elmegreen1977},
or trigger star formation in a pre-existing cloud \citep[e.g. \emph{radiatively driven implosion},][]{kessel2003}.

While the effect of photoevaporation has been observed using direct imaging \citep[{so-called \em{proplyd objects}},][]{odell1993,odell1994},
a proof of triggered star formation is harder to obtain.
Merely observing the presence of young stars in molecular clouds near hot stars is \emph{no solid evidence},
because it is not clear whether the birth of the stars was triggered by the hot star,
or whether stars were forming in these clouds regardless.
Moreover, photoevaporation may act to disperse the circumstellar environment of protostars as a function of their distance,
and therefore introduce an apparent evolutionary gradient which may incorrectly be interpreted as an age gradient 
\citep[e.g. discussed by][]{beltran2009}.

The relatively simple nature of IC\,1396, 
where the UV radiation field is dominated by the massive star system HD\,206267 (O6.5V), 
makes it a good region to study the influence of a hot star.
We note that the region also contains an O9V and several B-type stars \citep{simonson1968,garrison1976}, 
but these objects are thought to produce one (O9.5V) to five (B3V) orders of magnitude less ionising radiation compared to an O6.5V-type \citep{panagia1973,martins2005}.

In what follows, we discuss the hypothesis that star formation in IC\,1396 is triggered,
and test it by investigating the properties of our candidates as a function of their distance from HD\,206267.
We discover an age gradient which, we will argue, provides evidence for triggered star formation.

\subsubsection{Hypothesis: is star formation in IC\,1396 triggered?}
IC\,1396 is known to contain multiple dense molecular clouds \cite[e.g. BRC 32-42,][]{sugitani1991} 
containing a total mass between 9\,000 and 12\,000~\msol\ \citep{weikard1996,froebrich2005}.
The discovery of red IRAS point sources associated with these clouds 
first suggested that stars are actively forming here \citep{sugitani1991,schwartz1991}.
More recently, Spitzer and Chandra observations uncovered a large number 
of embedded protostars \cite[e.g][]{reach2004,sicilia2006spitzer,getman2007,ikeda2008,mercer2009,choudhury2010},
while optical studies have found Herbig Haro outflows \citep{reipurth1997,froebrich2005}.
These results indicate the presence of very young YSO-type objects \citep[$<$1~Myr,][]{evans2009} in the region,
although none of them are part of our sample because they are not detected in the optical.

Using CO radial velocities, \cite{patel1995} found that the clouds in the region are part of an expanding ring of molecular material 
centred roughly around HD\,206267 (dashed ellipse in Fig.~\ref{fig:spatial}). 
In their picture, the massive star was part of a first population which formed 3 to 4~Myr ago 
when the parent molecular cloud was compressed and flattened by turbulence (e.g., a supernova in the surrounding Cepheus OB2 association). 
Subsequently, wind or radiation from HD\,206267 caused leftover molecular material to be 
swept up in the expanding ring that is observed today,
and is thought to be the site of a second generation of star formation.

This picture is consistent with our results,
because many of our youngest objects are clustered 
just in front of the bright-rimmed clouds A, B and E (see Fig.~\ref{fig:spatial}).
The position of these objects between the clouds -- known to contain protostars -- 
and the massive star, 
suggests that they may have formed at an earlier stage when the shock front 
was located closer to the hot star.
The typical radial velocity dispersion in the region ($\sim$1-2~pc\,Myr$^{-1}$, 
SA06\footnote{\citet{sicilia2006} found a standard deviation of 3.6~km\,s$^{-1}$ (= 3.7~pc\,Myr$^{-1}$) in the radial velocity, 
but reported that individual errors contributed 1-2~km\,s$^{-1}$.})
suggests that our candidates may indeed still be located close to the place where they once formed.

\subsubsection{Test: does our sample show an age gradient?}
\label{sec:photoevaporation}
\begin{figure}
\includegraphics[width=\linewidth]{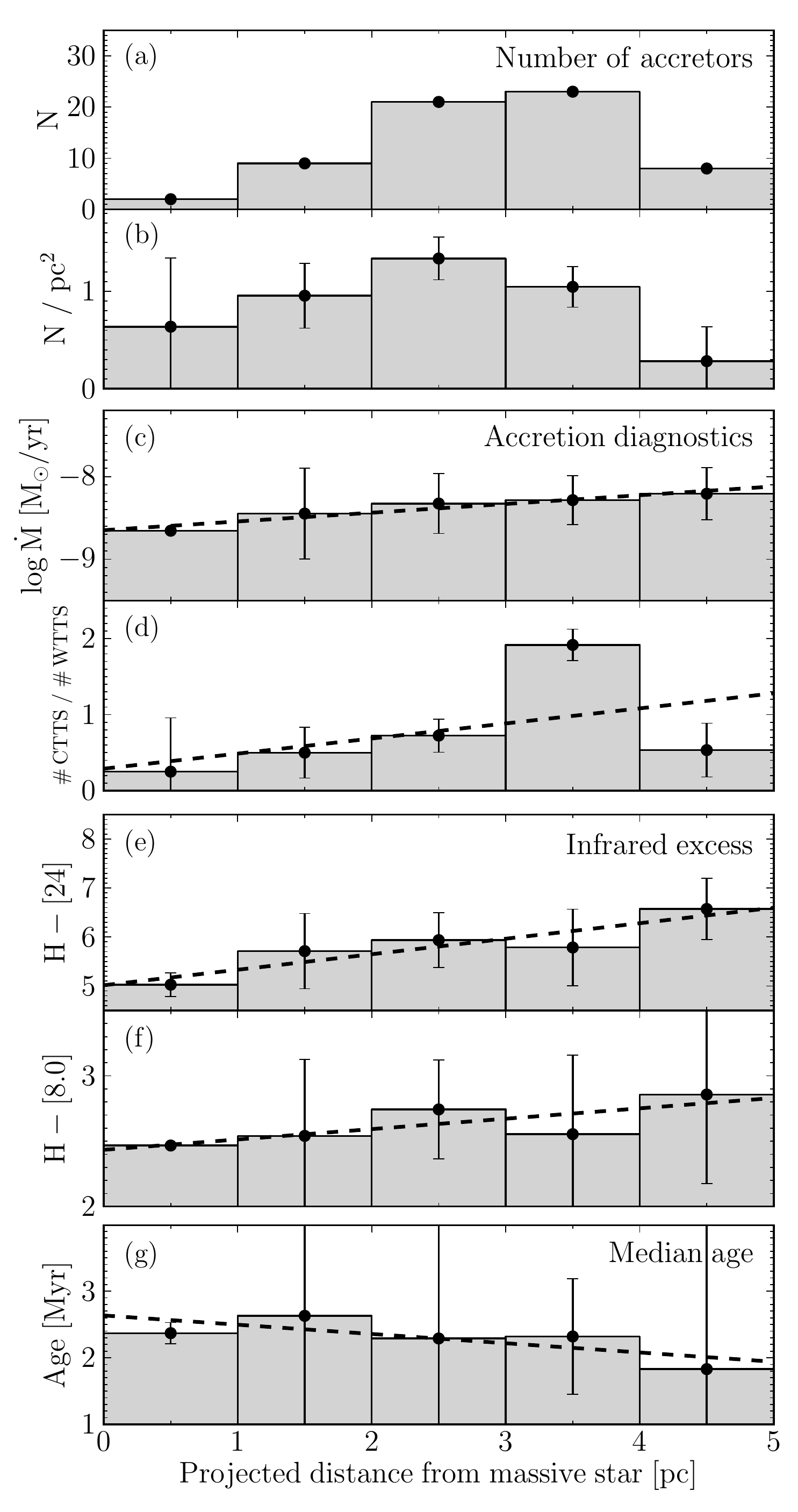}
\caption{\label{fig:radial} 
Candidate properties in function of the projected distance from HD\,206267,
limited out to 5~pc where both Spitzer and literature data are available.
With increasing distance from the massive star, we find higher mean accretion rates (panel c),
a higher fraction of accretors to the number of non-accretors from SA06 (panel d), 
stronger mean infrared disc excess (panels e-f),
and a lower median age (panel g).
The dashed lines show linear fits.
All panels are consistent with a spatio-temporal gradient where objects further away from the massive star have a younger appearance.
The bright rim of molecular cloud A, the Elephant Trunk Nebula, is located at the distance of 3.8~pc.
}
\end{figure}

Whether the clusters of stars observed in our sample have been triggered is a testable hypothesis.
If star formation is triggered, we expect that:
\begin{enumerate}
 \item the stars near the shock fronts must be younger than HD\,206267;
 \item there should be a small age spread and gradient; 
 stars located further away from HD\,206267 and closer to the shock fronts should be younger.
\end{enumerate}

Condition (i) is consistent with the fact 
that the model ages for the majority of our candidates ($\sim$2-3~Myr)
do not exceed the dynamical age of the expanding ring of molecular material \citep[2-3~Myr,][]{patel1995}.
While there is no reliable estimate of the age of HD\,206267 itself,
we note that the main-sequence lifetime of an O6.5V-type star \citep[4~Myr,][]{weidner2010} 
does not contradict the picture.

Proposed evidence for condition (ii), the age gradient, comes from 
the spatial distribution of the youngest members.
First, SA05 noted an increasing number of objects younger than 1~Myr towards globule A.
Subsequently, Spitzer observations of clouds A, E and J \citep{sicilia2006spitzer,getman2007,ikeda2008,choudhury2010}
showed that the embedded Class 0/I protostars are systematically located further away from HD\,206267 
than the more evolved optical Class II/III objects. 

We investigate if such gradient is also apparent in our sample.
In Fig.~\ref{fig:radial} we show the number of candidates as a function of the projected distance from HD\,206267 (panel~a)
and the corresponding surface density (panel~b).
We find a significant peak near $\sim$3~pc, 
which corresponds to cluster in front of cloud~A (commonly known as the Elephant's Trunk Nebula).
We also show the mean accretion rate (panel~c) 
and the number of weak-line (non-accreting) T-Tauri stars in the region obtained from SA06 (panel~d). 
We find an apparent increase in the accretion activity further away from the hot star.
This trend is consistent with an increase in the mean mid- and far-infrared excess (panels~e-f)
and a decrease in the median stellar age (panel~g).

Dashed lines show a linear regression of the mean properties as a function of radial distance.
Although the statistical significance of the individual correlations is only moderate at best 
(the p-values being 0.01, 0.4, 0.03, 0.12, 0.14 for panels c, d, e, f, g),
it is significant that all panels are mutually consistent with the picture of a spatio-temporal gradient towards cloud~A.

\subsubsection{Can the age gradient be explained by photoevaporation instead?}
Photoevaporation denotes the process where radiation heats disc particles 
until they reach the escape velocity and ``evaporate'' into space.
The effect is often discussed in the context of irradiation from the central star 
\citep[``inside-out'' photoevaporation, e.g.][]{ercolano2009},
but here we consider it in the context of the external radiation from HD\,206267 (``outside-in'').
\citet{beltran2009} argued that the infrared excess gradients observed in IC\,1396
would also be expected if such photoevaporation is progressively removing circumstellar material.
The Spitzer-based discovery by \citet{balog2006} of a tail on a proplyd-like object in the region, 
confirms that the effect certainly plays a role here.

However, the proplyd in question is located at only 0.5~pc from HD\,206267, 
while clouds A, B and E are located significantly further away at $\sim$4, 9, and 12~pc.
The available Spitzer data does not show proplyd-like tails around the stars in front of these clouds,
and SA05 did not find a clear absence of accreting stars near the O-star.
Both models and observations \citep{richling1998,balog2007} suggest that photoevaporation
is perhaps only effective at removing disks within $\sim$1\,pc from the radiation source.

Moreover, while a decrease of the infrared excess may be explained by photoevaporation,
it is not obvious that the removal of circumstellar material
from the outer disk would necessarily result in the lower mass accretion rates 
that are observed in Fig.~\ref{fig:radial}.
More importantly, it is hard to see how the apparent trend of decreasing model ages 
may be explained by photoevaporation 
(although the trend has a low statistical significance).

\begin{figure*}
\includegraphics[width=\linewidth]{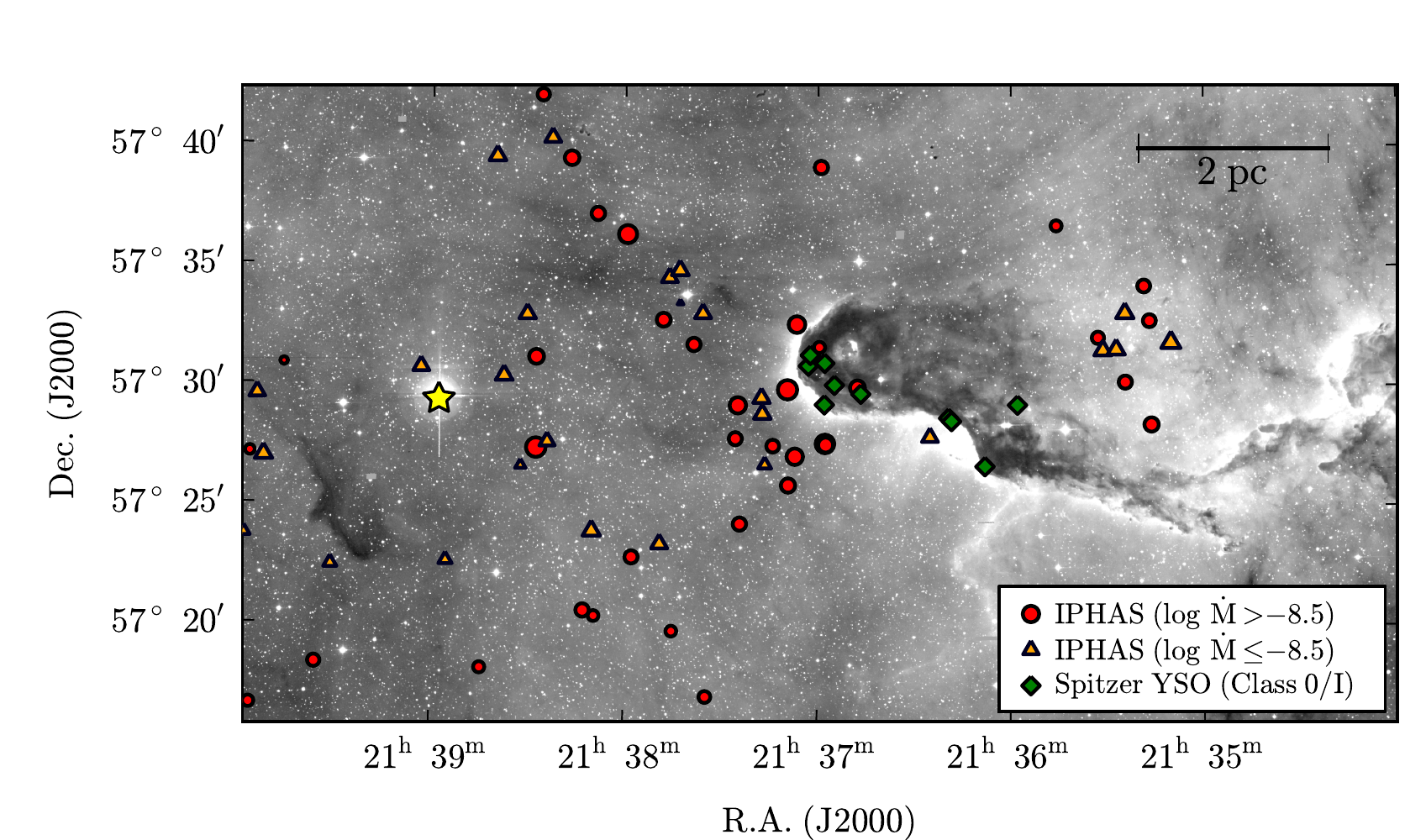}
\includegraphics[width=\linewidth]{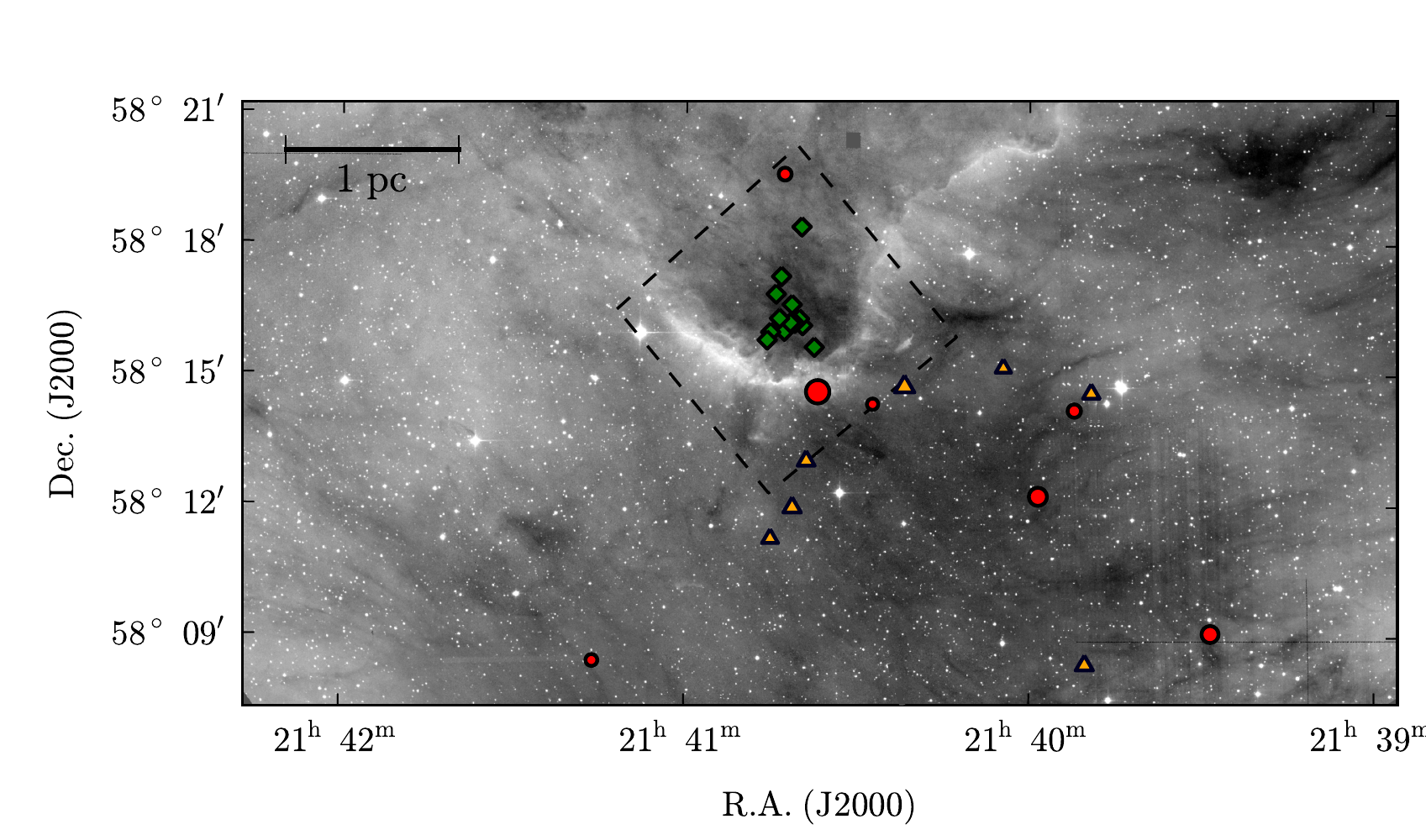}
\caption{\label{fig:closeup} 
Spatial distribution of our candidates (red circles and blue triangles) near the 
bright-rimmed clouds A/B (top panel) and cloud E (bottom panel).
The size of the symbol denotes the age.
We also show the position of Spitzer Class 0/I protostars (green diamonds) as 
found by \protect\citet[][top panel]{reach2009}
and \protect\citet[][bottom panel]{choudhury2010}.
The dashed line in the bottom panel shows the footprint of the Spitzer data.
HD\,206267 is indicated by a yellow star symbol in the top panel,
while it is located 12~pc south of the image in the bottom panel.
The dispersed nature of our candidates in front of the ionized rims, 
compared to the compact clustering of the Spitzer protostars inside the clouds,
suggests that our candidates were triggered at an earlier stage 
when the ionization shock was located closer to HD\,206267.}
\end{figure*}

The strongest argument in favour of triggered star formation 
comes from the dispersal of objects in front of clouds A/B/E, 
shown close-up in Fig.~\ref{fig:closeup}.
The clusters of our candidates in front of the ionized rims are significantly 
more dispersed than the embedded Class 0/I protostars inside the clouds,
covering an area which is roughly 1 to 2~pc larger in diameter.
If we were to assume that all objects in the region disperse from their birth 
at the rate of the cluster's current spread in radial velocity ($\sim$1-2 pc\,Myr$^{-1}$, see earlier), 
there would be a dynamical argument to suggest that our Class II candidates 
are more dispersed because they formed roughly 0.5 Myr before the Class 0/I protostars,
which is consistent with the apparent age gradient found in Fig.~\ref{fig:radial}.

In summary, we found evidence for a spatio-temporal gradient which is unlikely 
to be explained by photoevaporation alone.
This provides a strong indication that star formation in IC\,1396 has, at least in part, 
been triggered sequentially by HD\,206267 during the last $\sim$1~Myr.

\section{Conclusions}
We presented a photometric study of IC\,1396, located at a distance of 870~pc in the Cepheus OB2 association \citep{contreras}.
We used IPHAS narrow-band \halpha\ imaging and broadband \sloanr/\sloani\ photometry
to identify 158 low-mass T Tauri candidates from a database of 1 million objects,
using a homogeneous selection method which offers a precise handle on the selection effects.

Compared to previous searches by \citet{sicilia2005,sicilia2006}, 
we have tripled the number of classical T Tauri candidates in the region, 
in particular at low masses, with 56 per cent of our candidates below 0.5\,\msol.
Our main findings are:
\begin{enumerate}
 \item  We find a near-linear dependency of the accretion rate on stellar mass of $\rm \dot M_{acc} \propto M_\star^{1.1\pm0.2}$,
but the determination of the slope is affected by detectability limits and the limited mass range (0.2 - 2.0~\msol).
 \item A significant fraction of the candidates are found to be part 
 of dispersed clusters which are located in-between the 
 bright-rimmed molecular globules A/B/E and the massive star system HD\,206267 (O6.5V). 
We discovered a spatio-temporal gradient of increasing accretion rates, 
increasing disc excesses, and younger ages away from HD\,206267 towards globule A.
This provides a strong indication that the formation of these clusters has been 
sequentially triggered by the massive star \citep[in support of the picture proposed by][]{patel1995}.
This finding is consistent with recent Spitzer-based discoveries of apparent age gradients 
inside the globules \citep{getman2007,ikeda2008,choudhury2010}.
\end{enumerate}
In future work, we aim to extend our sample by applying the method introduced in this paper 
to different regions in the Galactic Plane, 
allowing comparative studies between different star-forming environments to be 
made based on a uniform selection method and large sample sizes.

A relaxed version of the selection criteria used in this work may also 
serve as a basis for high-resolution spectroscopic follow-up work.

\section*{Acknowledgments}
We thank the anonymous referee for providing useful comments which helped to improve the paper.
We are also indebted to A. Sicilia-Aguilar for providing helpful and constructive suggestions. 
Armagh Observatory is funded by major grants from the Department of Culture, Arts and Leisure (DCAL) for Northern Ireland 
and the UK Science and Technology Facilities Council (STFC).
The IPHAS survey is carried out at the Isaac Newton Telescope (INT). 
The INT is operated on the island of La Palma by the Isaac Newton Group in the Spanish Observatorio del Roque de los Muchachos of the Instituto de Astrofisica de Canarias. All IPHAS data are processed by the Cambridge Astronomical Survey Unit, at the Institute of Astronomy in Cambridge.
Our work made extensive use of the TopCat \citep{topcat}, Python, PyFITS and PostgreSQL software tools for data processing.
The mosaics were generated using the Montage software maintained by NASA/IPAC.

\label{lastpage}

\bsp

\appendix
\onecolumn

\begin{table}
\section{Simulated IPHAS colours for emission-line stars}
\label{app:tracks}
\caption{
Synthetic tracks in the $(r-H\alpha,r'-i')$ plane 
for objects with increasing levels of \halpha\ emission equivalent width (EW).
The underlying data are provided by the library of SEDs due to \citet{pickles},
assuming a standard extinction law $\rm R_{V} = 3.1$ (see \S \ref{sec:colours}).}
\centering
\begin{tabular}{rrlrlrlrlrl} \hline
\halpha & \multicolumn{10}{c}{Reddening} \\
EW(\AA) & \multicolumn{2}{c}{$E(B-V)=0.0$} &  \multicolumn{2}{c}{$E(B-V)=1.0$} &  \multicolumn{2}{c}{$E(B-V)=2.0$} &  \multicolumn{2}{c}{$E(B-V)=3.0$} &  \multicolumn{2}{c}{$E(B-V)=4.0$} \\
 & $(r' - i')$ & $(r'-H\alpha)$ & $(r' - i')$ & $(r'-H\alpha)$ & $(r' - i')$ & $(r'-H\alpha)$ & $(r' - i')$ & $(r'-H\alpha)$ & $(r' - i')$ & $(r'-H\alpha)$ \\ \hline
O5V SED & & & & & & & & & & \\
0	 & -0.179 & 0.076    & 0.509 & 0.302    & 1.182 & 0.486		& 1.841 & 0.629	& 2.486 & 0.733 \\
-10	 & -0.185 & 0.174    & 0.502 & 0.399    & 1.174 & 0.582		& 1.831 & 0.724	& 2.476 & 0.827 \\
-20	 & -0.190 & 0.262    & 0.495 & 0.486    & 1.166 & 0.668		& 1.822 & 0.809	& 2.466 & 0.911 \\
-40	 & -0.201 & 0.419    & 0.482 & 0.640    & 1.150 & 0.819		& 1.804 & 0.958	& 2.446 & 1.059 \\
\hline
\end{tabular}
\label{tab:tracks}
\end{table}

\begin{table*}
\caption{Table A1 -- continued for M-type SEDs.}
\centering
\begin{tabular}{rrlrlrlrlrl} \hline
\halpha & \multicolumn{10}{c}{Reddening} \\
EW(\AA) & \multicolumn{2}{c}{$E(B-V)=0.0$} &  \multicolumn{2}{c}{$E(B-V)=1.0$} &  \multicolumn{2}{c}{$E(B-V)=2.0$} &  \multicolumn{2}{c}{$E(B-V)=3.0$} &  \multicolumn{2}{c}{$E(B-V)=4.0$} \\
 & $(r' - i')$ & $(r'-H\alpha)$ & $(r' - i')$ & $(r'-H\alpha)$ & $(r' - i')$ & $(r'-H\alpha)$ & $(r' - i')$ & $(r'-H\alpha)$ & $(r' - i')$ & $(r'-H\alpha)$ \\ \hline
M0V SED & & & & & & & & & & \\
0	 & 0.929 & 0.537    & 1.599 & 0.699    & 2.255 & 0.821		& 2.898 & 0.906	& 3.530 & 0.958 \\
-10	 & 0.920 & 0.635    & 1.589 & 0.796    & 2.243 & 0.917		& 2.885 & 1.001	& 3.517 & 1.053 \\
-20	 & 0.911 & 0.723    & 1.579 & 0.883    & 2.232 & 1.003		& 2.873 & 1.086	& 3.504 & 1.137 \\
-40	 & 0.894 & 0.878    & 1.559 & 1.035    & 2.210 & 1.153		& 2.849 & 1.234	& 3.479 & 1.284 \\
\hline
\end{tabular}
\label{tab:tracks2}
\end{table*}
\twocolumn

\LTcapwidth=0.9\textwidth

\onecolumn
\section{Photometry of T Tauri candidates towards IC\,1396}
\label{app:photometry}
\scriptsize
\begin{longtable}{rcccccccc}
\caption{\label{tab:candidates} Photometry of T-Tauri candidates towards IC\,1396,
selected based on their position in the IPHAS $(r-H\alpha,r'-i')$ and $(r',r'-i')$ diagrams.
The IAU-registered naming convention for the objects is formed 
by prefixing ``IPHAS'' to the position string given in column 2.
Calibrated IPHAS photometry is given in columns 3-5
and matched 2MASS data is given in columns 6-8.
For objects which were previously identified in the survey
by \citet{sicilia2006} we give the existing identifier in column 9.
} \\
\hline
  & IPHAS ID         & \multicolumn{3}{c}{IPHAS photometry} & \multicolumn{3}{c}{2MASS magnitudes} & Name \\ 
  & J[RA(2000)$+$Dec(2000)] & $r'$ & $r'-i'$ & $r'-H\alpha $ & $J$ & $H$ & $K$ & [SA06]\tablenotemark{1} \\
\hline
\endfirsthead

\multicolumn{3}{c}%
{{\bfseries \tablename\ \thetable{} -- continued}} \\
\hline
  & IPHAS ID          & \multicolumn{3}{c}{IPHAS photometry} & \multicolumn{3}{c}{2MASS magnitudes} & Name \\ 
  & J[RA(2000)$+$Dec(2000)] & $r'$ & $r'-i'$ & $r'-H\alpha $ & $J$ & $H$ & $K$ & [SA06]\tablenotemark{1} \\
\hline
\endhead

\hline \hline
\endlastfoot
\footnotetext[1]{[SA06]: identifier from \citet{sicilia2006}.} 
  1 & J213008.57+575727.4 &  19.93$\pm$0.03 & 2.04$\pm$0.04 & 1.40$\pm$0.04  & 15.50$\pm$0.08 & 14.85$\pm$0.08 & 14.53$\pm$0.09 &  \\
  2 & J213237.11+583241.6 &  17.85$\pm$0.01 & 1.29$\pm$0.02 & 1.45$\pm$0.02  & 14.56$\pm$0.03 & 13.80$\pm$0.04 & 13.57$\pm$0.04 &  \\
  3 & J213317.77+574813.5 &  14.73$\pm$0.00 & 0.68$\pm$0.00 & 0.96$\pm$0.00  & 12.25$\pm$0.02 & 10.96$\pm$0.03 & 9.83$\pm$0.02 &  \\
  4 & J213324.81+575022.9 &  17.33$\pm$0.00 & 1.01$\pm$0.01 & 0.96$\pm$0.01  & 14.53$\pm$0.03 & 14.02$\pm$0.05 & 13.65$\pm$0.04 &  \\
  5 & J213450.88+583534.7 &  19.61$\pm$0.03 & 1.58$\pm$0.04 & 1.66$\pm$0.04  & 15.74$\pm$0.10 & 14.85$\pm$0.11 & 14.42$\pm$0.09 &  \\
  6 & J213451.70+575140.5 &  18.33$\pm$0.01 & 1.58$\pm$0.02 & 1.28$\pm$0.02  & 14.33$\pm$0.03 & 13.25$\pm$0.03 & 12.65$\pm$0.04 &  \\
  7 & J213453.08+575125.2 &  18.73$\pm$0.01 & 1.52$\pm$0.02 & 1.35$\pm$0.02  & 14.71$\pm$0.04 & 13.48$\pm$0.04 & 12.68$\pm$0.03 &  \\
  8 & J213510.14+565022.6 &  18.04$\pm$0.01 & 1.18$\pm$0.01 & 0.88$\pm$0.02  & 15.15$\pm$0.05 & 14.47$\pm$0.06 & 13.92$\pm$0.05 &  \\
\end{longtable}
\normalsize
\twocolumn

\onecolumn
\section{Derived parameters of IPHAS T Tauri candidates towards IC\,1396}
\label{app:properties}
\scriptsize
\begin{longtable}{rccccccc}
\caption{\label{tab:properties} 
Derived parameters of the T Tauri candidates found towards IC\,1396.
Estimates for the age, mass and radius are based on the \citet{siess} 
model tracks placed in the $(r',r'-i')$ diagram (Fig.~\ref{fig:cmd}).
The \halpha\ equivalent width (EW) is derived from the $(r-H\alpha,r'-i')$  
diagram and was subsequently used to obtain the absolute line excess luminosity $L(\halpha)$ 
and a corresponding mass accretion rate $\dot M$ (see \S \ref{sec:analysis}).
The parameters were obtained using the assumption that 
(i) all stars are placed at a distance of 870~pc,
(ii) all stars have a reddening $A_V = 1.56$,
and (iii) there is a direct relation between the \halpha\ luminosity and the accretion luminosity.
The estimated $1\sigma$-uncertainties that follow from these assumptions are the errors listed
in this table, obtained using Monte Carlo simulations (see \S \ref{sec:errors}).
In the case of the age, the 25\% and 75\% quartiles are given as error bounds.
} \\
\hline
 & IPHAS ID       	&   Age   & Mass & Radius	& EW($\halpha$)	&  $\rm \log L(\halpha)$	&  $\rm \log \dot M$ \\ 
 & J[RA(2000)$+$Dec(2000)] & [Myr] & [$\rm M_\odot$] & [$\rm R_\odot$]	& [\AA] 	& [$\rm L_\odot$]		&  [$\rm M_\odot\,yr^{-1}$]     \\
\hline
\endfirsthead

\multicolumn{3}{c}%
{{\bfseries \tablename\ \thetable{} -- continued}} \\
\hline
 & IPHAS ID     & Age   & Mass & Radius	& EW($\halpha$)	&  $\rm \log L(\halpha)$	&  $\rm \log \dot M$ \\ 
 & J[RA(2000)$+$Dec(2000)]    & [Myr] & [$\rm M_\odot$] & [$\rm R_\odot$]	& [\AA] 	& [$\rm L_\odot$]		&  [$\rm M_\odot\,yr^{-1}$]     \\
\hline
\endhead

\hline \hline
\endlastfoot

  1 & J213008.57+575727.4 &  $ 4.0^{-0.5}_{+0.5}$ & $0.28 \pm 0.04$ & $0.91 \pm 0.17$ &  $\   -52 \pm   7$  & $-3.8 \pm 0.2$ & $-9.2 \pm 0.6$ \\
  2 & J213237.11+583241.6 &  $ 4.0^{-0.5}_{+0.4}$ & $0.53 \pm 0.10$ & $1.13 \pm 0.10$ &  $ -114 \pm   7$  & $-2.7 \pm 0.2$ & $-8.2 \pm 0.6$ \\
  3 & J213317.77+574813.5 &  $ 12.9^{-6.8}_{+4.7}$ & $1.46 \pm 0.09$ & $1.69 \pm 0.17$ &  $\   -81 \pm   6$  & $-1.8 \pm 0.2$ & $-7.4 \pm 0.6$ \\
  4 & J213324.81+575022.9 &  $ 18.7^{-1.5}_{+0.9}$ & $0.85 \pm 0.12$ & $0.98 \pm 0.08$ &  $\   -52 \pm   5$  & $-2.9 \pm 0.2$ & $-8.7 \pm 0.6$ \\
  5 & J213450.88+583534.7 &  $ 9.2^{-1.2}_{+1.5}$ & $0.34 \pm 0.05$ & $0.71 \pm 0.11$ &  $ -136 \pm  11$  & $-3.3 \pm 0.2$ & $-8.9 \pm 0.6$ \\
  6 & J213451.70+575140.5 &  $ 2.6^{-0.3}_{+0.4}$ & $0.41 \pm 0.05$ & $1.15 \pm 0.14$ &  $\   -60 \pm   5$  & $-3.1 \pm 0.2$ & $-8.6 \pm 0.6$ \\
  7 & J213453.08+575125.2 &  $ 4.9^{-0.5}_{+0.7}$ & $0.41 \pm 0.05$ & $0.94 \pm 0.13$ &  $\   -76 \pm   6$  & $-3.2 \pm 0.2$ & $-8.7 \pm 0.6$ \\
  8 & J213510.14+565022.6 &  $ 17.1^{-3.3}_{+2.6}$ & $0.68 \pm 0.12$ & $0.86 \pm 0.08$ &  $\   -31 \pm   4$  & $-3.4 \pm 0.2$ & $-9.2 \pm 0.6$ \\
\end{longtable}
\normalsize
\twocolumn

\onecolumn
\section{Spitzer IRAC and MIPS photometry}
\label{app:spitzer}
\scriptsize
\begin{longtable}{cccccc}
\caption{\label{tab:spitzer} Spitzer IRAC and MIPS magnitudes for the T Tauri 
candidates in the central region (Spitzer data does not cover the entire IC\,1396 region). 
The uncertainties listed include the background noise, 
but not the $\sim$10 per cent uncertainty in the absolute calibration which is typical for Spitzer.
See \S \ref{sec:infrared} for details on the data reduction. 
} \\
\hline
IPHAS ID        	&   3.6 \micron   & 4.5 \micron & 5.8 \micron	& 8 \micron	&  24 \micron	\\ 
\hline
\endfirsthead

\multicolumn{3}{c}%
{{\bfseries \tablename\ \thetable{} -- continued}} \\
\hline
IPHAS ID        	&   3.6 \micron   & 4.5 \micron & 5.8 \micron	& 8 \micron	&  24 \micron	\\ 
\hline
\endhead

\hline \hline
\endlastfoot

J213531.36+573127.8 & 12.396 $\pm$ 0.003 & 12.151 $\pm$ 0.002 & 12.037 $\pm$ 0.008 & 11.731 $\pm$ 0.008 & 8.465 $\pm$ 0.037  \\
J213532.89+573159.6 & 13.030 $\pm$ 0.002 & 12.653 $\pm$ 0.002 & 12.138 $\pm$ 0.006 & 11.119 $\pm$ 0.005 & 8.310 $\pm$ 0.032  \\
J213625.07+572750.2 & 12.430 $\pm$ 0.002 & 11.862 $\pm$ 0.003 & 11.395 $\pm$ 0.004 & 10.346 $\pm$ 0.003 & 7.537 $\pm$ 0.016  \\
J213659.09+573905.6 & 11.619 $\pm$ 0.002 & 11.151 $\pm$ 0.001 & 10.562 $\pm$ 0.002 & 9.482 $\pm$ 0.001 & 5.455 $\pm$ 0.003  \\
J213659.47+573134.8 & 11.699 $\pm$ 0.002 & 11.269 $\pm$ 0.001 & 10.765 $\pm$ 0.003 & 9.598 $\pm$ 0.003  & - \\
J213706.50+573231.8 & 10.803 $\pm$ 0.002 & 10.369 $\pm$ 0.001 & 9.984 $\pm$ 0.002 & 9.386 $\pm$ 0.001 & 6.901 $\pm$ 0.010  \\
J213707.03+572700.8 & 10.684 $\pm$ 0.002 & 10.388 $\pm$ 0.001 & 10.165 $\pm$ 0.002 & 9.310 $\pm$ 0.001 & 5.833 $\pm$ 0.003  \\
J213709.09+572548.6 & 13.057 $\pm$ 0.002 & 12.972 $\pm$ 0.002 & 12.829 $\pm$ 0.011 & 12.536 $\pm$ 0.015 & 9.395 $\pm$ 0.085  \\
\end{longtable}
\normalsize
\twocolumn

\onecolumn
\section{Rejected candidates}
\label{app:rejected}
\scriptsize
\begin{longtable}{rcccccccc}
\caption{\label{tab:rejected} IPHAS and 2MASS photometry of objects towards IC\,1396
which, based on their outlier position in the $(r-H\alpha,r'-i')$ diagram,
are likely to be \halpha\ emission-line objects.
However, their position in the $(r',r'-i')$ suggests that 
they are not related to IC\,1396 itself,
so they have not been considered as T Tauri candidates in our work.
See \S \ref{sec:selection} for details.} \\
\hline
  & IPHAS ID         & \multicolumn{3}{c}{IPHAS photometry} & \multicolumn{3}{c}{2MASS magnitudes}  \\ 
  & J[RA(2000)$+$Dec(2000)] & $r'$ & $r'-i'$ & $r'-H\alpha $ & $J$ & $H$ & $K$  \\
\hline
\endfirsthead

\multicolumn{3}{c}%
{{\bfseries \tablename\ \thetable{} -- continued}} \\
\hline
  & IPHAS ID          & \multicolumn{3}{c}{IPHAS photometry} & \multicolumn{3}{c}{2MASS magnitudes}  \\ 
  & J[RA(2000)$+$Dec(2000)] & $r'$ & $r'-i'$ & $r'-H\alpha $ & $J$ & $H$ & $K$  \\
\hline
\endhead

\hline \hline
\endlastfoot
  1 & J213412.45+562519.6 &  17.83$\pm$0.01 & 0.05$\pm$0.02 & 0.44$\pm$0.01  &       -      &      -      &      -        &  \\
  2 & J213415.13+564734.3 &  18.91$\pm$0.02 & 1.14$\pm$0.03 & 1.65$\pm$0.02  & 15.61$\pm$0.07 & 14.66$\pm$0.08 & 13.79$\pm$0.06 &  \\
  3 & J213454.21+562406.8 &  15.16$\pm$0.00 & 0.33$\pm$0.00 & 0.41$\pm$0.00  & 13.23$\pm$0.02 & 12.95$\pm$0.04 & 12.86$\pm$0.03 &  \\
  4 & J213459.66+580325.9 &  18.02$\pm$0.01 & 0.86$\pm$0.01 & 0.69$\pm$0.02  & 15.62$\pm$0.09 & 15.43$\pm$0.16 & 15.05$\pm$0.14 &  \\
  5 & J213502.78+572747.1 &  16.79$\pm$0.00 & 0.68$\pm$0.01 & 0.57$\pm$0.01  & 14.89$\pm$0.03 & 14.67$\pm$0.07 & 14.36$\pm$0.09 &  \\
  6 & J213520.74+561049.3 &  16.60$\pm$0.00 & 0.02$\pm$0.01 & 0.20$\pm$0.01  & 16.41$\pm$0.14 & $>$ 16.49    & $>$ 16.19    &  \\
  7 & J213557.95+572532.2 &  18.50$\pm$0.01 & 0.88$\pm$0.02 & 0.77$\pm$0.02  & 16.11$\pm$0.11 & 15.49$\pm$0.14 & 15.42$\pm$0.19 &  \\
  8 & J213626.91+570213.7 &  15.86$\pm$0.00 & 0.65$\pm$0.00 & 0.61$\pm$0.00  & 14.15$\pm$0.03 & 13.78$\pm$0.04 & 13.56$\pm$0.04 &  \\
\end{longtable}
\normalsize
\twocolumn

\end{document}